\documentclass[reprint,amsmath,amssymb,aps,superscriptaddress,floatfix,prmaterials]{revtex4-2}

\usepackage{graphicx}
\usepackage{dcolumn}
\usepackage{bm}
\usepackage{color}
\usepackage[caption=false]{subfig}

\newcommand*\LCC{Li$_2$CoCl$_4$}

\begin{document}

\title{Zero-field magnetic structure and metamagnetic phase transitions of the cobalt chain compound Li$_2$CoCl$_4$}

\author{Zachary W. Riedel}
\affiliation{Department of Materials Science and Engineering, University of Illinois at Urbana-Champaign, Urbana, Illinois 61801, United States}
\affiliation{Materials Research Laboratory, University of Illinois at Urbana-Champaign, Urbana, Illinois 61801, United States}

\author{Zhihao Jiang}
\affiliation{Department of Materials Science and Engineering, University of Illinois at Urbana-Champaign, Urbana, Illinois 61801, United States}

\author{Maxim Avdeev}
\affiliation{Australian Nuclear Science and Technology Organisation, New Illawarra Rd, Lucas Heights, NSW 2234, Australia}
\affiliation{School of Chemistry, The University of Sydney, Sydney, NSW 2006, Australia}

\author{Andr\'e Schleife}
\affiliation{Department of Materials Science and Engineering, University of Illinois at Urbana-Champaign, Urbana, Illinois 61801, United States}
\affiliation{Materials Research Laboratory, University of Illinois at Urbana-Champaign, Urbana, Illinois 61801, United States}
\affiliation{National Center for Supercomputing Applications, University of Illinois at Urbana-Champaign, Urbana, IL 61801, United States}

\author{Daniel P. Shoemaker}
\email{dpshoema@illinois.edu}
\affiliation{Department of Materials Science and Engineering, University of Illinois at Urbana-Champaign, Urbana, Illinois 61801, United States}
\affiliation{Materials Research Laboratory, University of Illinois at Urbana-Champaign, Urbana, Illinois 61801, United States}

\begin{abstract}
Exploring the uncharacterized magnetic phases of Co$^{2+}$ chain compounds is critical for finding new low-dimensional magnets hosting quantized excitations. We map the unexplored magnetic phases of the Co$^{2+}$ chain compound \LCC. Magnetometry reveals magnetic ordering below 7~K with a metamagnetic transition near 16.5~kOe and a gradual transition to a field-aligned paramagnetic state above 31~kOe. Curie-Weiss fits to the high temperature susceptibility reveal a high-spin (\mbox{spin-$\frac{3}{2}$}) state for cobalt. Heat capacity data, though, give a magnetic entropy change of 5.46~J/mol, consistent with cobalt effective \mbox{spin-$\frac{1}{2}$} systems. To characterize the zero-field antiferromagnetic ordering, we separately calculated the energy of proposed magnetic structures with density functional theory and collected 3.5~K neutron diffraction data, finding that \LCC\ has ferromagnetic chains with antiferromagnetic interactions between them. Increasing field rotates these spin chains, producing the antiferromagnetic to intermediate to paramagnetic transition sequence.
\end{abstract}

\maketitle

\section{Introduction}
Low-dimensional \mbox{spin-$\frac{1}{2}$} magnets are ideal systems for probing quantum ground and excited states, providing opportunities for realizing Ising and Heisenberg magnetism models and for hosting quantized excitations. Co$^{2+}$ in octahedra, notably, can have a low-spin configuration (\mbox{spin-$\frac{1}{2}$}, t$_{2g}^{6}$e$_{g}^{1}$). Co$^{2+}$ also can have a high-spin configuration (\mbox{spin-$\frac{3}{2}$}, t$_{2g}^{5}$e$_{g}^{2}$) where spin-orbit coupling and polyhedral distortions lead to a ground state Kramers doublet (\textit{J}=$\frac{1}{2}$) that is energetically well-separated from higher energy states and can dominate low temperature behavior,\cite{goodenough1968spin,piwowarska2019origin,winter2022magnetic} effectively producing \mbox{spin-$\frac{1}{2}$} properties.\cite{ogawa1965specific,ortega2005factors,shirata2012experimental,thota2021magnetic,zhang2022magnetic} Mapping the magnetic phase regions of unexplored octahedral Co$^{2+}$ compounds is thus necessary to assess their applicability for studying low-dimensional \mbox{spin-$\frac{1}{2}$} behavior.

Of interest here are alkali-Co$^{2+}$-halide compounds, which often contain one-dimensional cobalt chains. High-spin CsCoBr$_3$, CsCoCl$_3$, and RbCoCl$_3$ have anti-aligned magnetic moments along their cobalt chains and experimental data matches Ising antiferromagnet behavior,\cite{yelon1975magnetic,mekata1978magnetic,melamud1974magnetic,hanni2021magnetic} with CsCoCl$_3$ and CsCoBr$_3$ experimental data also matching effective \mbox{spin-$\frac{1}{2}$} behavior.\cite{achiwa1969linear,buyers1986solitons} As a result, these materials have been used to study quantum spin excitations.\cite{braun2007chiral,mena2020thermal} Another material with anti-aligned intrachain moments, Cs$_2$CoCl$_4$, has Co$^{2+}$ cations arranged in cobalt-chlorine tetrahedra, and it has an effective \mbox{spin-$\frac{1}{2}$} state\cite{kenzelmann2002order} that has encouraged its use in studying quantum phase transitions\cite{mukherjee2004field} and in developing an entanglement detection protocol.\cite{laurell2021quantifying} 

While searching for similar one-dimensional magnetic compounds using a magnetic dimensionality toolkit,\cite{karigerasi2018uncovering} we found \LCC. The compound is another alkali-cobalt-halide with Co$^{2+}$ chains and with octahedral coordination. It has a low temperature phase with \textit{Cmmm} space group symmetry,\cite{schneider1993polymorphie} and near 300$^{\circ}$C, differential thermal analysis,\cite{urbanowicz1985dta} differential scanning calorimetry, and neutron powder diffraction have shown a transition to a disordered rock-salt structure.\cite{schneider1993polymorphie} The low temperature phase has nearly-regular, edge-sharing cobalt-chlorine octahedra that form chains parallel to the unit cell's \textit{c}-axis. Studied only for its electronic transport properties,\cite{wussow1989lattice,gupta1997lattice} \LCC\ has unexplored magnetic properties. This structure, promisingly, is similar to that of high-spin CoCl$_{2}\cdot$2H$_2$O, which has field-dependent magnetic phases and experimental results matching effective \mbox{spin-$\frac{1}{2}$} Ising chain behavior.\cite{kobayashi1964metamagnetic,narath1964antiferromagnetism,narath1964magnetization,shinoda1964heat,cox1966neutron,weitzel1974neutron,yamakawa1979temperature} Moreover, CoCl$_{2}\cdot$2D$_2$O has observed quantum criticality and spin excitations.\cite{montfrooij2001spin,larsen2017spin} Therefore, probing the magnetic structure of \LCC\ will provide further insights into the magnetic behavior of its cobalt chains and their potential for studying quantum phenomena.

\section{Materials and methods}
\subsection{Synthesis}
We ground LiCl (99.9\%, Alfa Aesar) and CoCl$_2$ (99.7\%, Alfa Aesar) together under argon in the stoichiometric ratio and placed the mixture in a quartz tube. Highly hygroscopic LiCl cannot be exposed to air or moisture. Therefore, we used an Edwards Speedivalve to close the tube under argon and transfer it to a vacuum pump line. The vacuum line was then pumped at 45~mTorr for 15~min to clear it before opening the valve to pump argon from the sample tube. We heated the sealed mixture at 10$^{\circ}$C/min to 550$^{\circ}$C, held for 12~h, and then cooled to room temperature at 10$^{\circ}$C/min. Samples were royal blue and solidified in chunks, having been heated above the melting temperature.

\subsection{Characterization}
We used a Bruker D8 Advance equipped with a capillary geometry to collect X-ray diffraction (XRD) data with Mo-K$\alpha$ radiation. We sealed XRD samples under vacuum in a thin glass capillary. This process involved momentary exposure to air.  

Zero-field neutron diffraction data were collected on $\sim$2~g powder at 3.5~K and 15~K on the ECHIDNA beamline\cite{avdeev2018echidna} at the Australian Centre for Neutron Scattering using neutrons with a 2.4395~\AA\ wavelength. We solved the magnetic structure using \textsc{FullProf}\cite{rodriguez1993fullprof} and \textsc{GSAS-II},\cite{toby2013gsas} giving the same result. The \textsc{GSAS-II} magnetic structure solution additionally used the Bilbao Crystallographic Server's \textsc{k-Subgroupsmag} program.\cite{perez2015symmetry} Unit cell images were produced with \textsc{VESTA}.\cite{momma2011vesta}

We collected susceptibility and magnetization data with a Quantum Design Magnetic Property Measurement System (MPMS3) on powder samples. For 2 to 400~K data sets collected at 8, 16, 25, and 45~kOe, we used the vibrating sample magnetometry (VSM) mode. For magnetic phase mapping, which involved fields up to 70~kOe, we used the DC mode since the MPMS3 VSM fitting was poor at the highest moment values. We measured the compound's zero-field-cooled and field-cooled susceptibility from 2.5 to 15~K at 10, 100, and 1000~Oe before stepping in 5~kOe intervals from 5 to 70~kOe. We also collected isothermal magnetization curves from -70 to 70~kOe at 2.5~K and at 3 to 9~K at 1~K intervals. 

For heat capacity data, we used a Quantum Design Physical Property Measurement System. We attached a 1.1~mg chunk of polycrystalline \LCC\ to a calibrated puck with Apiezon N-grease and subtracted the grease contribution. We measured the heat capacity from 3 to 145~K at zero field and from 3 to 40~K for applied fields up to 55~kOe.

\subsection{Computational procedure}
We performed spin-polarized density functional theory (DFT) calculations with the Vienna \textit{Ab-initio} Simulation Package (\textsc{VASP}) for 12 proposed magnetic configurations. Total energies for each were calculated using the Perdew-Burke-Ernzerhof (PBE) generalized gradient approximation (GGA) to describe the exchange-correlation energy.\cite{perdew1996generalized} We used the cell parameters and atomic positions from our 3.5~K neutron diffraction experiment as a starting point for relaxations.\cite{supplement} Only the atomic geometry was provided, not the magnetic structure, to avoid biasing our calculations. 

Convergence tests found that a plane-wave cutoff of 800~eV with a 3$\times$2$\times$6 $\Gamma$-centered \mbox{\textbf{k}-point} grid for single unit cell magnetic configurations converged energy and pressure values well. For configurations requiring $c$-axis doubling, a 3$\times$2$\times$3 grid was used. We relaxed the atomic positions, cell shape, and cell volume of the 3.5~K cell with a force tolerance of 5~meV/\AA, performing a collinear calculation and initializing the cobalt magnetic moments to $\pm$1~${\mu}_{\mathrm{B}}$. Then we performed a non-collinear calculation with spin-orbit coupling. Band structure and density of states calculations for the lowest energy configuration were performed using the SCAN \mbox{meta-GGA} functional without spin-orbit coupling.\cite{sun2016accurate}

\section{Results and discussion}
\subsection{Phase purity}
\begin{figure}
    \centering
    \includegraphics[width=\columnwidth]{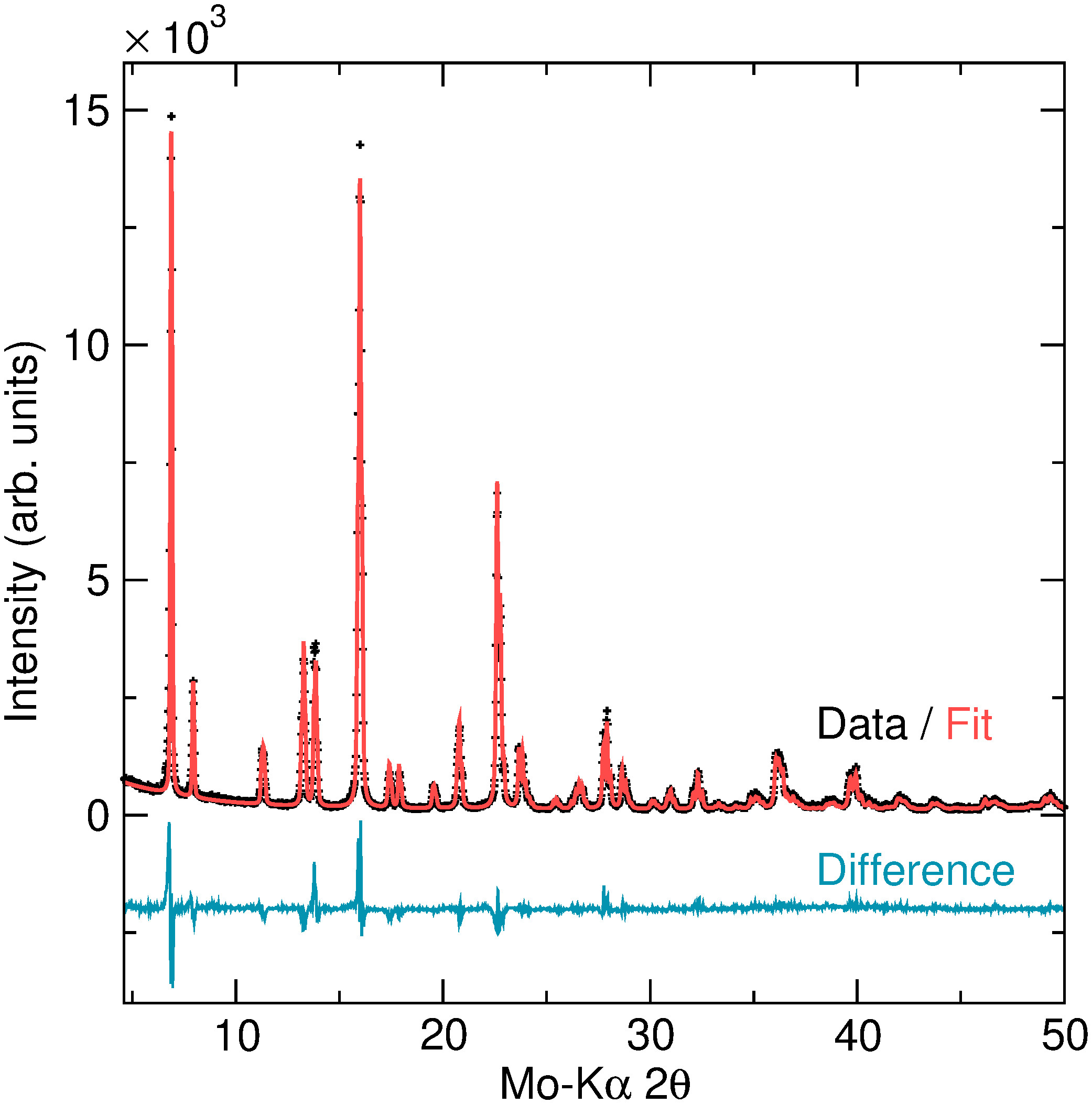}
    \caption{Room temperature powder XRD data matches the expected low temperature \textit{Cmmm} space group phase.}
    \label{fig:XRD_nuclear-cell}
\end{figure}

\begin{figure}
    \centering
    \includegraphics[width=0.85\columnwidth]{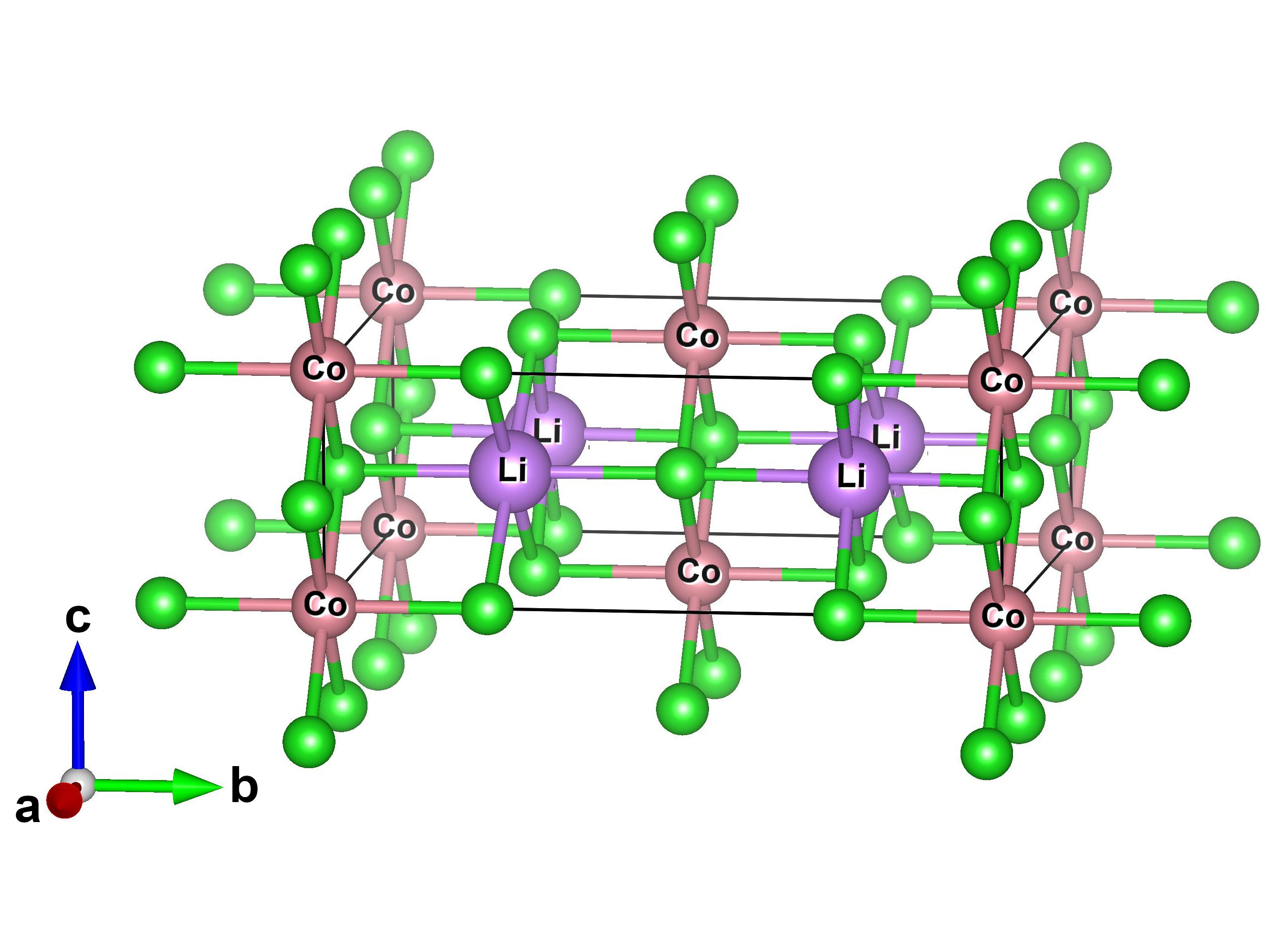}
    \caption{The low temperature phase unit cell contains cobalt-chlorine octahedra, forming chains along the \textit{c}-axis. Similarly, lithium-chlorine octahedra chains run along the \textit{a}-axis.}
    \label{fig:nuclear_cell}
\end{figure}

Powder XRD confirmed sample phase purity (Fig.~\ref{fig:XRD_nuclear-cell}) with Rietveld refinements matching the low temperature \textit{Cmmm} polymorph (Fig.~\ref{fig:nuclear_cell}). If \LCC\ is exposed to air, the XRD pattern shows unidentified impurity peaks within minutes, so keeping the samples in an inert environment is critical. Note that though \LCC\ was originally reported to crystallize in the \textit{Immm} space group,\cite{kanno1987new,kanno1988structure} this was corrected.\cite{schneider1993polymorphie} 

The compound has an order-disorder phase transition, determined previously by differential thermal analysis and differential scanning calorimetry, between 270$^{\circ}$C and 316$^{\circ}$C.\cite{urbanowicz1985dta,kanno1987new,schneider1993polymorphie} We collected \textit{in-situ} XRD data up to 480$^\circ$C, which showed the transition near 330$^\circ$C. Upon cooling, the compound transformed back into the low temperature phase at the same temperature.\cite{supplement} We could not stabilize the high temperature phase at room temperature by quenching 430$^{\circ}$C powder in ice water; quenching resulted in a pure low temperature phase product. The high temperature phase, therefore, prevents simple single crystal growth by slow cooling the melt. After confirming phase purity, we performed basic magnetic measurements to map the magnetic phase diagram.

\subsection{Magnetic susceptibility and magnetization}
\begin{figure}
    \centering\includegraphics[width=0.95\columnwidth]{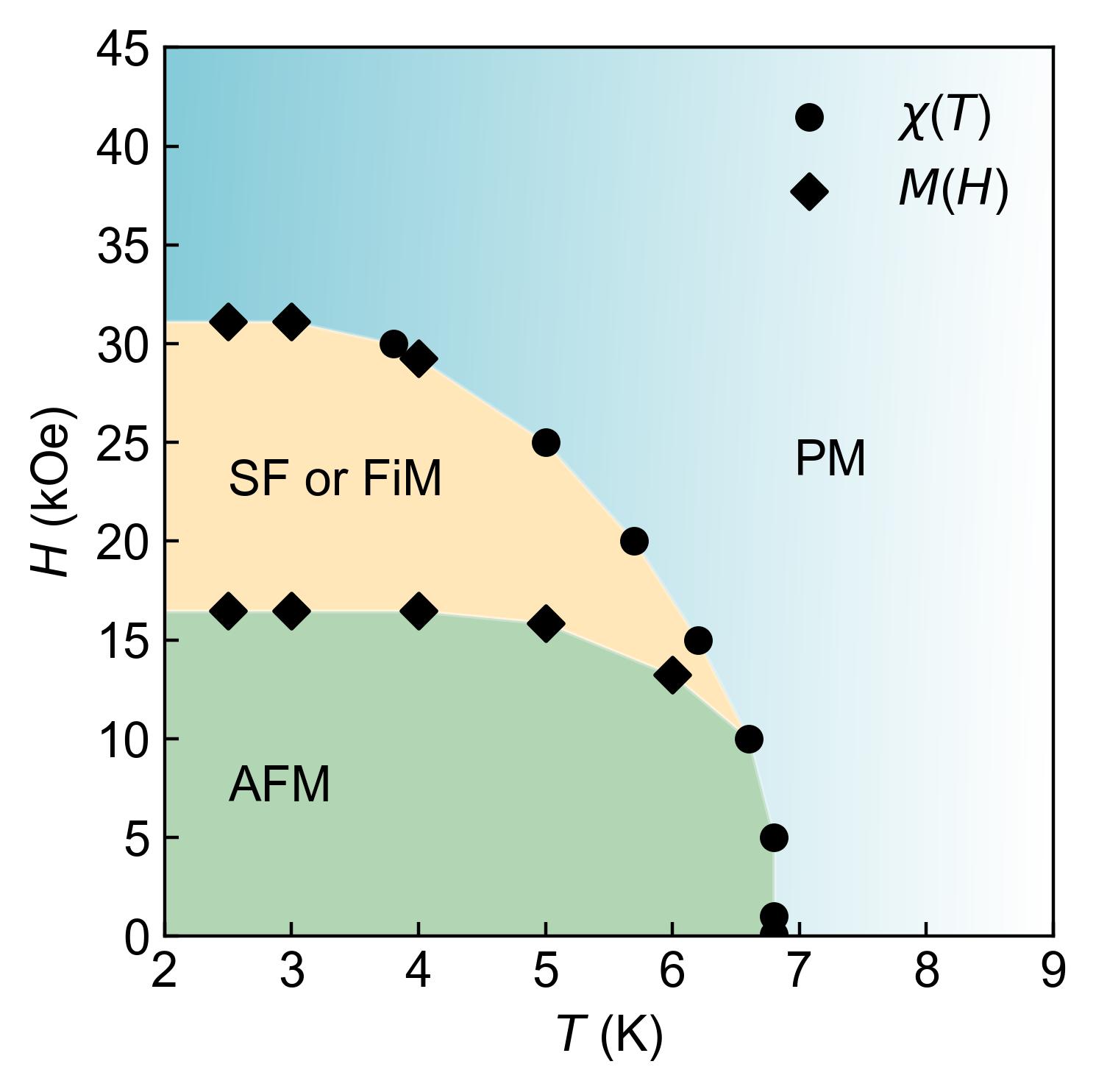}
    \caption{The magnetic phase regions of \LCC\ are derived from susceptibility transition temperatures and from isothermal magnetization curve derivatives. At 2.5~K, an antiferromagnetic phase transitions to an intermediate phase (spin-flop or possibly ferrimagnetic) around 16.5~kOe and from the intermediate phase to a nearly saturated paramagnetic phase around 31~kOe. The powder data indicates gradual changes during the field-driven transitions. Data point shapes correspond to their origin.}
    \label{fig:magPD}
\end{figure}

The magnetic susceptibility at low temperature shows a distinct peak and downturn, indicating antiferromagnetic (AFM) order. At 1~kOe, the N\'eel temperature (\textit{T}$_{\mathrm{N}}$) is 6.8~K. The transition peak broadens with increasing field until above 30~kOe the susceptibility no longer decreases at lower temperatures. As the field increases above 30~kOe, the moment begins to saturate, leading to a decrease in susceptibility. Moment saturation along with the absence of a sharp transition in the susceptibility indicates a paramagnetic state with field-induced moment alignment. Representative spectra are shown in Fig.~\ref{fig:magPDsets} with additional data in the Supplemental Material.\cite{supplement} In the magnetization curves, we observe no sharp transitions; rather, we see gradual transitions that we identify with derivative curve peaks (Fig.~\ref{fig:magPDsets}). It is, of course, unclear how sharp these transitions may be for single crystal measurements along specific axes. The derivative peaks indicate three magnetic ordering regions. Combining the peaks with susceptibility-derived transition temperatures gives the magnetic phase diagram boundaries in Fig.~\ref{fig:magPD}.

\begin{figure*}
    \centering
    \subfloat{\includegraphics[width=0.95\columnwidth]{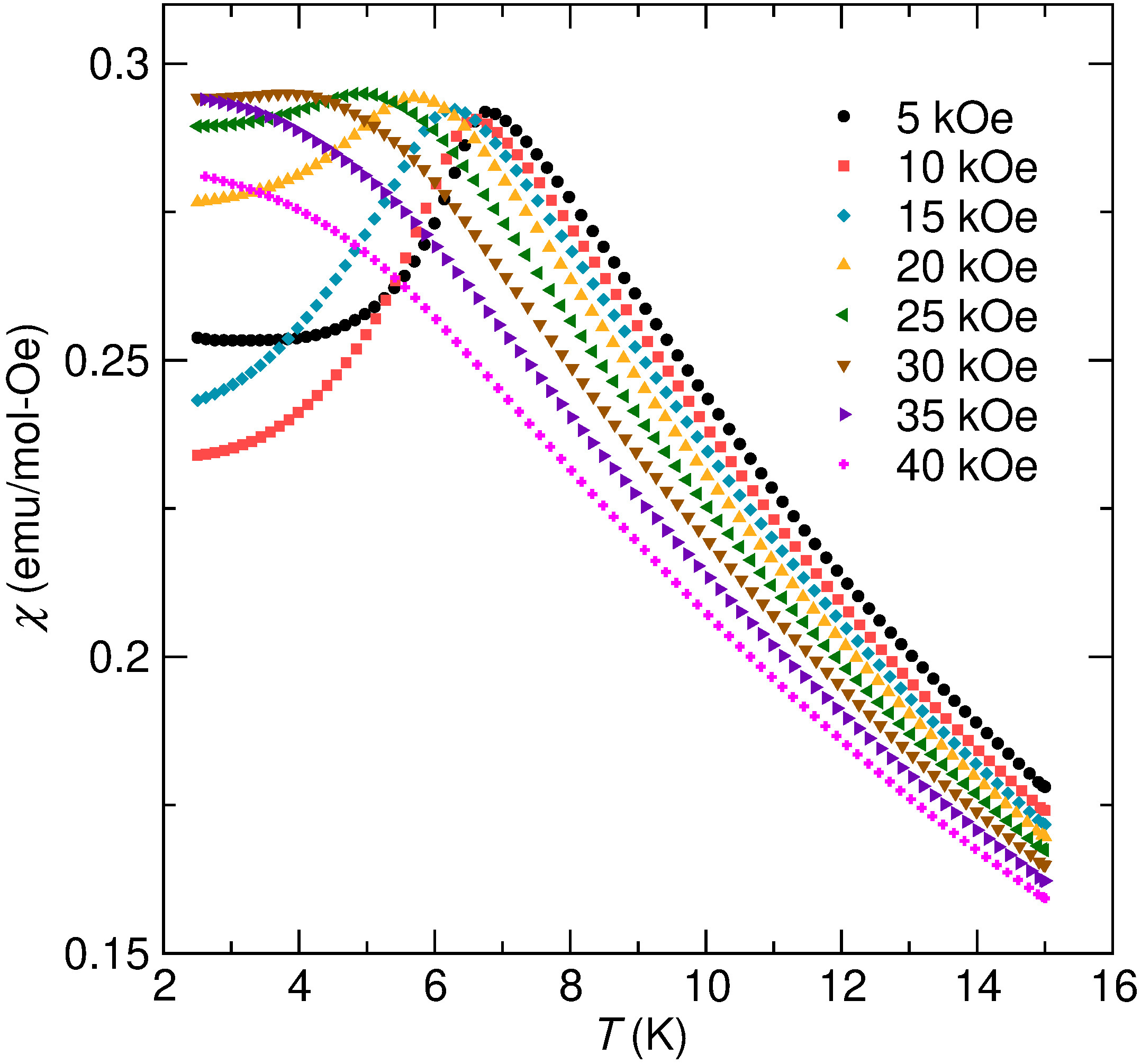}}
    \hfill
    \subfloat{\includegraphics[width=0.915\columnwidth]{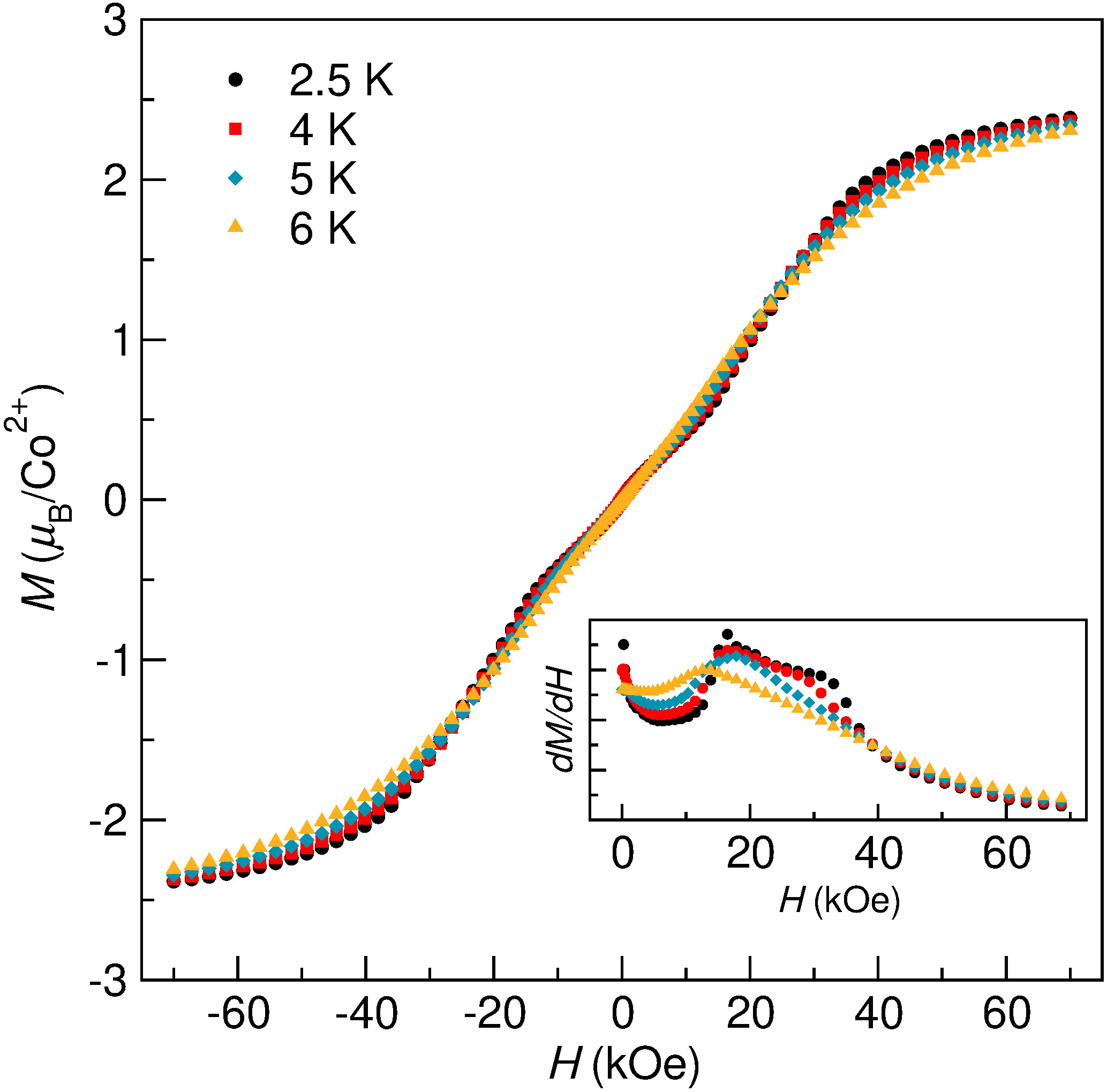}}
    \caption{(Left) Zero-field-cooled magnetic susceptibility data collected at 5$\leq$\textit{H}$\leq$40~kOe show observable transition temperatures used to construct the magnetic phase diagram. Data below 5~kOe had a tail at low temperature, presumably from impurity spins, that was suppressed at higher fields.\cite{supplement} Data above 30~kOe continues to decrease in susceptibility, indicating a saturated paramagnetic state.\cite{supplement} (Right) Magnetization curves show slope changes associated with metamagnetic phase transitions. The shifts appear as peaks in the derivative plot inset.}
    \label{fig:magPDsets}
\end{figure*}

The boundaries between the low field (AFM), intermediate field, and high field regions are not sharp. The magnetization derivative peaks separating the regions, likewise, broaden with increasing temperature. At 70~kOe and 2.5~K, the measured 2.38~$\mu_{\mathrm{B}}$/Co$^{2+}$ magnetization appears close to saturation, and at the proposed transition from the AFM phase to the intermediate field phase, the magnetization is 0.76~$\mu_{\mathrm{B}}$/Co$^{2+}$, roughly one-third the saturated moment. CoCl$_{2}\cdot$2H$_2$O, noted before because of its similar structure to \LCC, has an antiferromagnetic to ferrimagnetic (FiM) transition followed by a FiM to paramagnetic (PM) transition with increasing field.\cite{cox1966neutron,weitzel1974neutron,montfrooij2001spin} At low fields, CoCl$_{2}\cdot$2H$_2$O has ferromagnetic chains with antiferromagnetic interactions between them, and with increasing field, one-third of the moments rotate to form the FiM phase. Consequently, the FiM phase magnetization is one-third that of the saturated moment. The similar magnetization behavior of \LCC\ indicates that it may transition with increasing field from AFM to FiM to PM behavior. Our data, though, is also consistent with an intermediate field spin-flop (SF) phase.\cite{carlin1986magnetochemistry} Therefore, further neutron diffraction experiments with applied field are needed to confirm the intermediate field phase ordering. 

\begin{figure}
    \centering\includegraphics[width=\columnwidth]{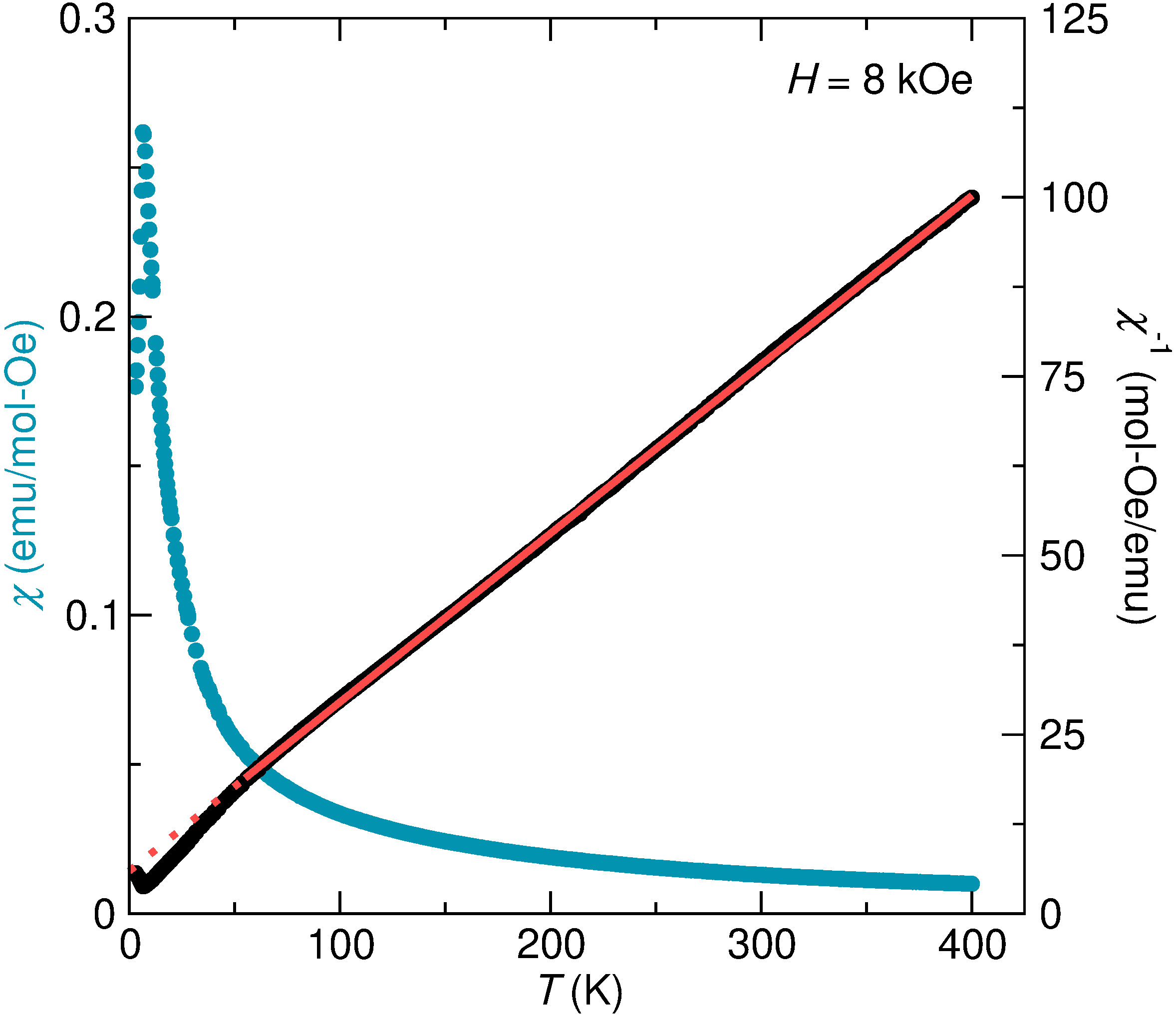}
    \caption{Field-cooled magnetic susceptibility data collected at 8~kOe follows Curie-Weiss paramagnetism above 55~K (R$^2$~=~0.99992) with $\theta_{\mathrm{CW}}$~=~\mbox{-25.3}~K indicating antiferromagnetic order.}
    \label{fig:MvT_8kOe}
\end{figure}

Fig.~\ref{fig:MvT_8kOe} shows a Curie-Weiss fit to susceptibility data from 55 to 400~K at 8~kOe. This gives $\theta_{\mathrm{CW}}$~=~\mbox{-25.3}~K and an effective moment of 5.83~$\mu_{\mathrm{B}}$, indicating dominant antiferromagnetic interactions and a large, unquenched orbital moment. The frustration index of 3.8 ($\lvert\theta_{\mathrm{CW}}$/$T_{\mathrm{N}}\rvert$=$\lvert$-25.3/6.7$\rvert$) also indicates mild frustration. The moment magnitude is consistent with other \mbox{spin-$\frac{3}{2}$}  Co$^{2+}$ compounds with regular or nearly regular octahedra (frequently double perovskites).\cite{martinez2002preparation,viola2003structure,ortega2005factors,ivanov2010structural} At higher fields, linear fits to data from 55 to 400~K also give negative Curie-Weiss parameters; however, deviations from linear behavior indicate competing ferromagnetic and antiferromagnetic interactions.\cite{supplement} Also, the inverse susceptibility shifts slightly below 55~K at all fields, no longer obeying Curie-Weiss paramagnetism. The shift is subtle, and we could not account for the curvature with a temperature-independent term. No anomalies appear in the zero-field heat capacity at that temperature, which suggests the onset of short-range magnetic correlations.  

\subsection{Heat capacity}\label{sec:HC}
\begin{figure}
    \centering\includegraphics[width=0.975\columnwidth]{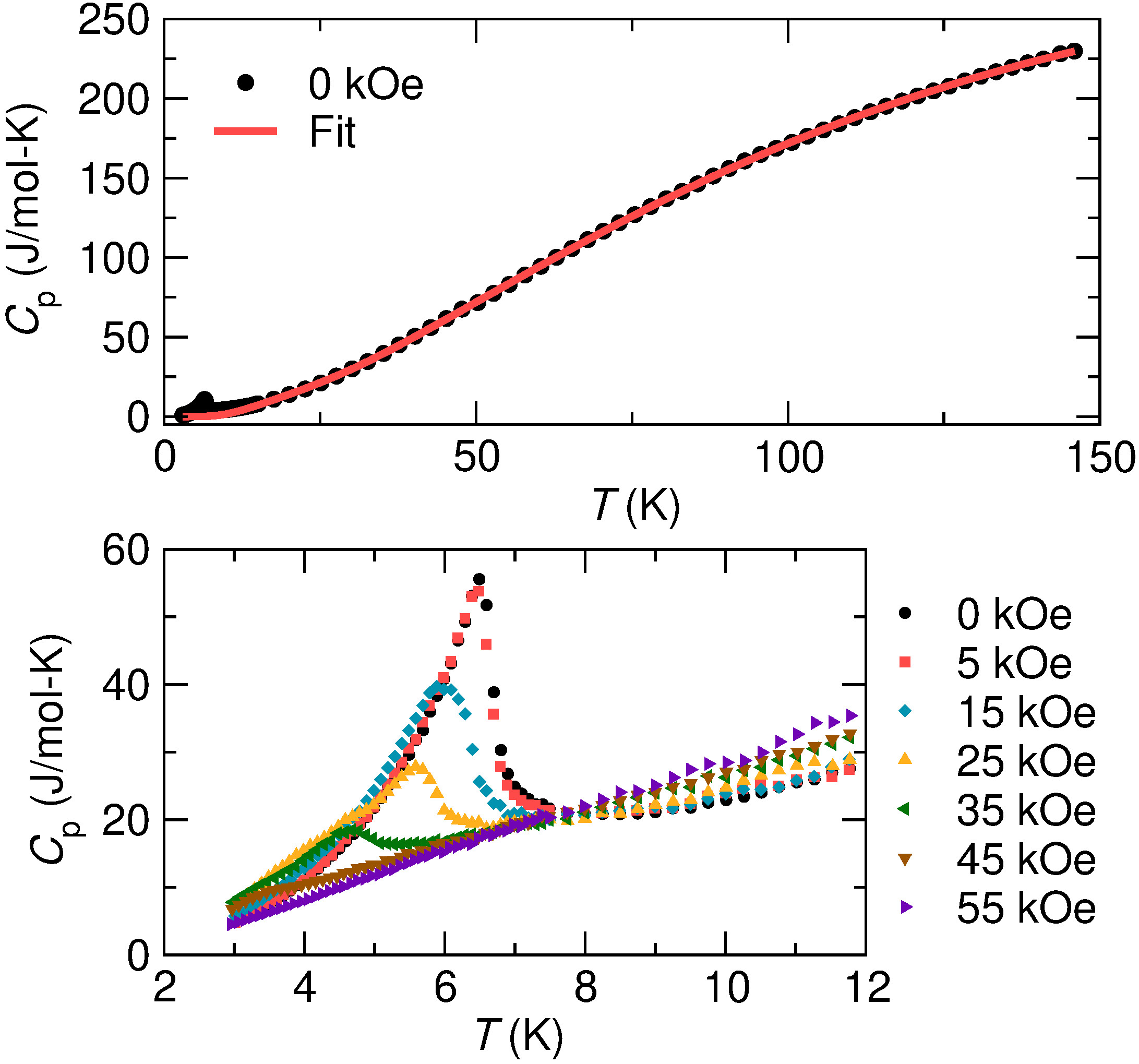}
    \caption{(Top) The zero-field heat capacity is plotted along with a Debye-Einstein fit to data above 15.5~K. The N\'eel temperature appears as a peak. (Bottom) Low temperature heat capacity data is plotted for increasing field. The zero-field and 5~kOe data are nearly identical. Above the intermediate phase to PM transition found with magnetometry, there remains a peak until around 55 kOe, where the low temperature heat capacity is linear.}
    \label{fig:ZFHC}
\end{figure}

Fig.~\ref{fig:ZFHC} contains the heat capacity data. The heat capacity is similar at zero field and 5~kOe, showing a transition at the N\'eel temperature. In the intermediate field region at 25~kOe, there is a small kink in the transition peak at 4.9~K, reflecting the destabilization of the antiferromagnetic behavior with increasing field. There is also a peak at 4.6~K with an applied field of 35~kOe and a small hump at 3.5~K with a 45~kOe field. 

To isolate the magnetic heat capacity contribution, we fit the lattice contribution of the zero-field data above 15.5~K to a Debye-Einstein equation with one Debye term and two Einstein terms (Fig.~\ref{fig:ZFHC}). Based on diffraction and published computational data,\cite{jain2013commentary,de2015charting} we estimated that the difference between \textit{C}$_\mathrm{p}$ and \textit{C}$_\mathrm{v}$ is negligible. Fit information is in the Supplemental Material,\cite{supplement} and Fig.~\ref{fig:magHC} shows the isolated magnetic heat capacity. At 25~kOe applied field, a hump in the magnetic heat capacity above the susceptibility transition temperature appears. This hump is characteristic of one-dimensional systems with short-range order.\cite{de2001experiments,sebastian2021quasi,ivanov2010structural,algra1976heat} At 55~kOe, only the hump remains. This corresponds with the high field susceptibility data where at low temperature the curve no longer decreases and appears paramagnetic with a saturated moment.\cite{supplement} Integrating \textit{C}$_\mathrm{p,mag}$/\textit{T} over temperature gives the magnetic entropy change during these transitions. For the zero-field data, the entropy change reaches 5.46~J~mol$^{-1}$ (Fig.~\ref{fig:magHC}). This is 94.7\% of the expected value of \textit{R}ln(2) for \mbox{spin-$\frac{1}{2}$} systems and suggests a Kramers doublet ground state with effective \mbox{spin-$\frac{1}{2}$} (\textit{J}=$\frac{1}{2}$), consistent with behavior seen in other high-spin cobalt materials.\cite{ortega2005factors,ogawa1965specific,algra1976heat}

\begin{figure}
    \centering\includegraphics[width=\columnwidth]{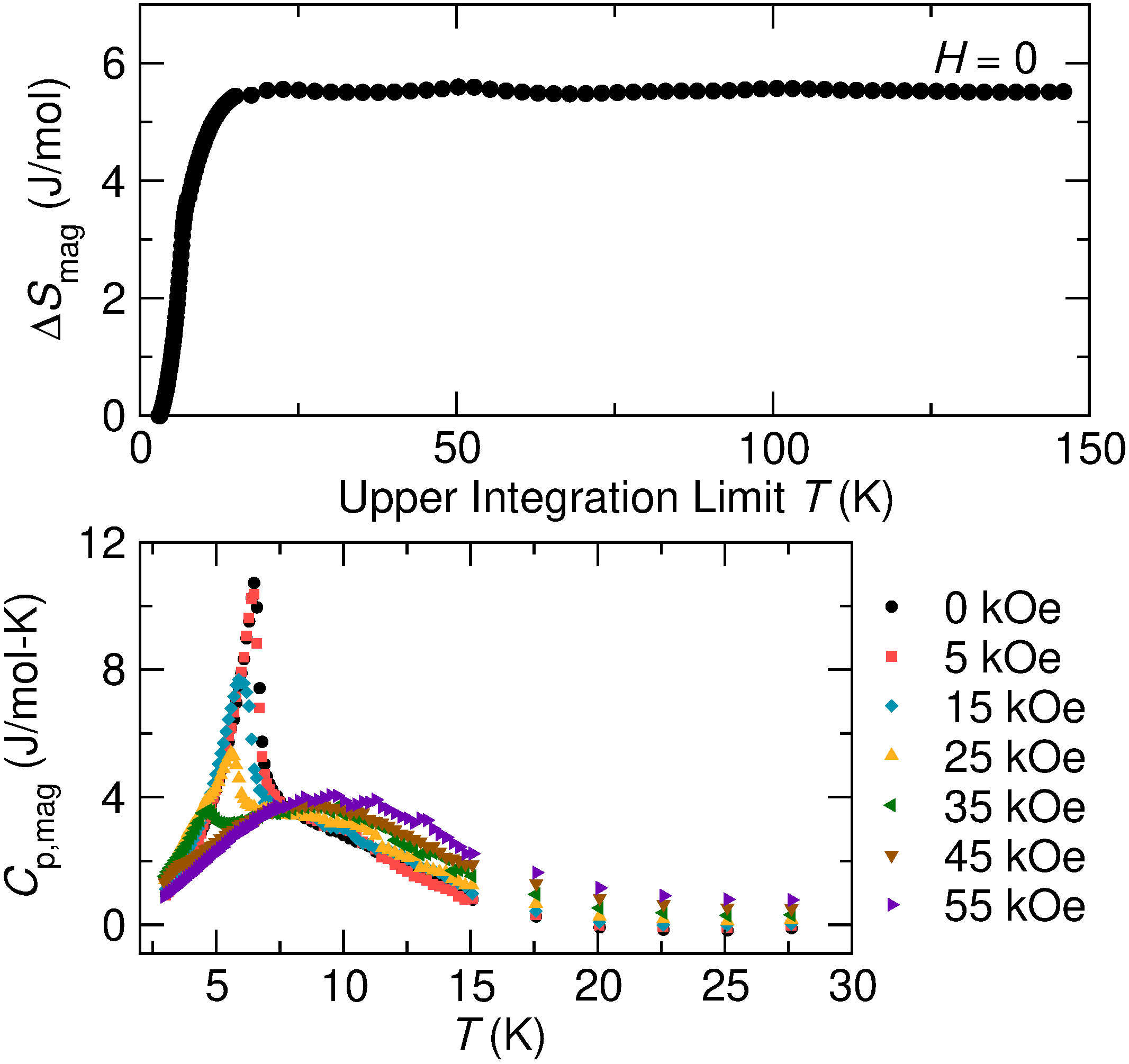}
    \caption{(Top) The magnetic entropy change is calculated over a range of integration intervals, showing a total change of 5.46 J/mol. (Bottom) The magnetic contribution to the heat capacity is plotted, showing a transition from long-range to short-range order with increasing field.}
    \label{fig:magHC}
\end{figure} 

\subsection{DFT magnetic structure and spin state}
To probe the magnetic ordering, we performed DFT total energy calculations on 12 potential magnetic configurations without consulting neutron diffraction results. Each configuration was defined by its cobalt intrachain and interchain interactions as well as its moment direction. A-C had ferromagnetic chains with antiferromagnetic interactions between them; D-F had antiferromagnetic chains with ferromagnetic interactions between them; G-I had antiferromagnetic chains with antiferromagnetic interactions between them; and J-L were ferromagnetic. Each set of interaction types included a configuration with moments along the $a$-axis (A, D, G, J), $b$-axis (B, E, H, K), and $c$-axis (C, F, I, L). Unit cells for all 12 are in the Supplemental Material.\cite{supplement}

\begin{figure}
    \centering\includegraphics[width=0.85\columnwidth]{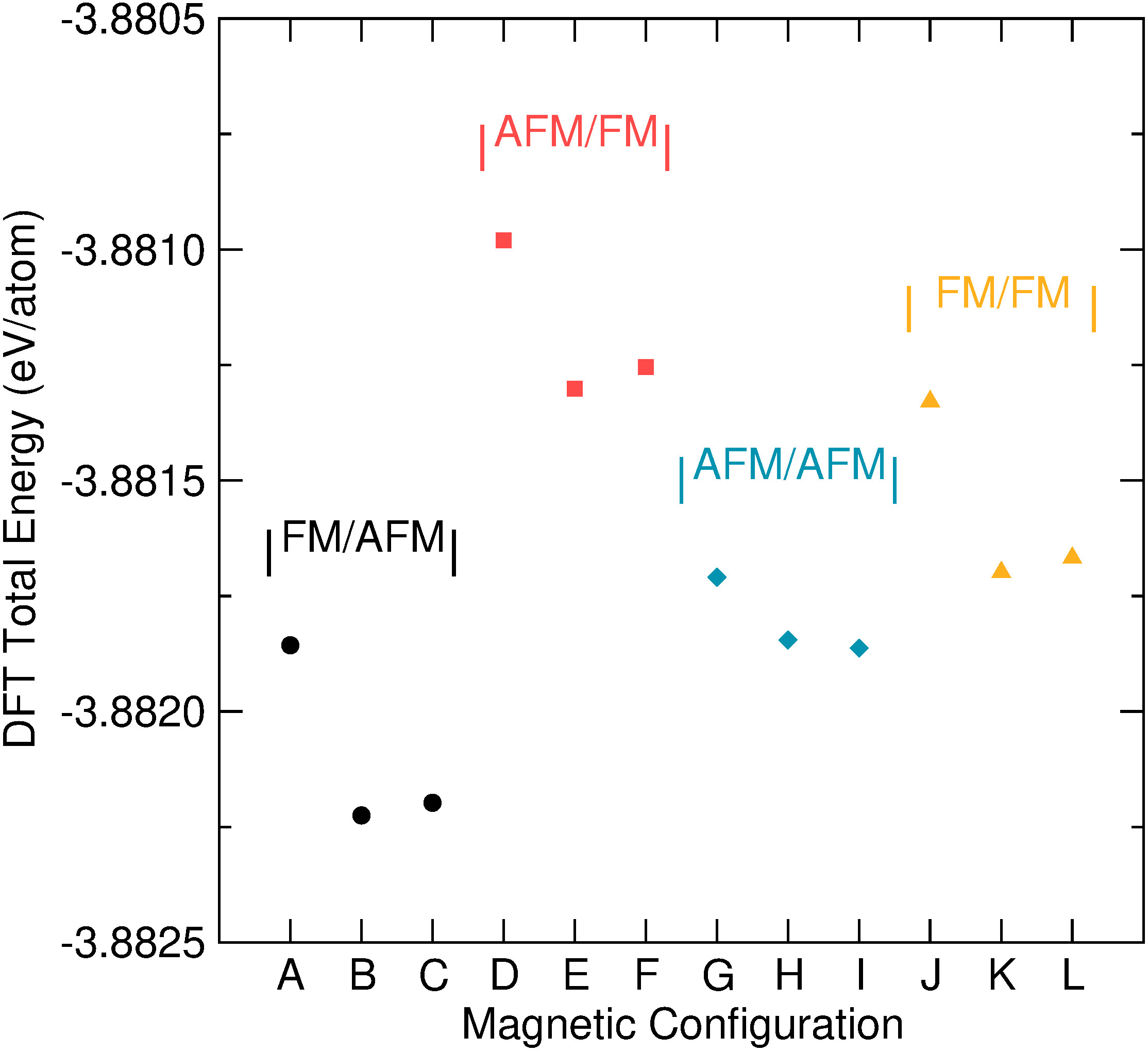}
    \caption{DFT-calculated total energies (GGA) are shown for each magnetic configuration. The three groups are labeled by their intra-/inter-chain interactions. Generally, FM/AFM configurations were the most stable and AFM/FM ones the least. Moments along the $a$-axis were considerably less favorable than along $b$ or $c$.}
    \label{fig:dftE}
\end{figure}

As expected, differences between the total energies are small, but trends are apparent (Fig.~\ref{fig:dftE}). For each set of interaction types, the configuration with moments along the $a$-axis had the highest energy, indicating that the $a$-axis should be the hard axis. The differences between $b$-axis and $c$-axis configurations, on the other hand, were much smaller. Regarding average energies for each interaction type set, the ferromagnetic chain with antiferromagnetic interactions between chains set (A-C) had the lowest energy while the inverse antiferromagnetic intrachain paired with ferromagnetic interchain interactions (D-F) had the highest. 

\begin{figure}
    \subfloat{\includegraphics[width=0.48\columnwidth]{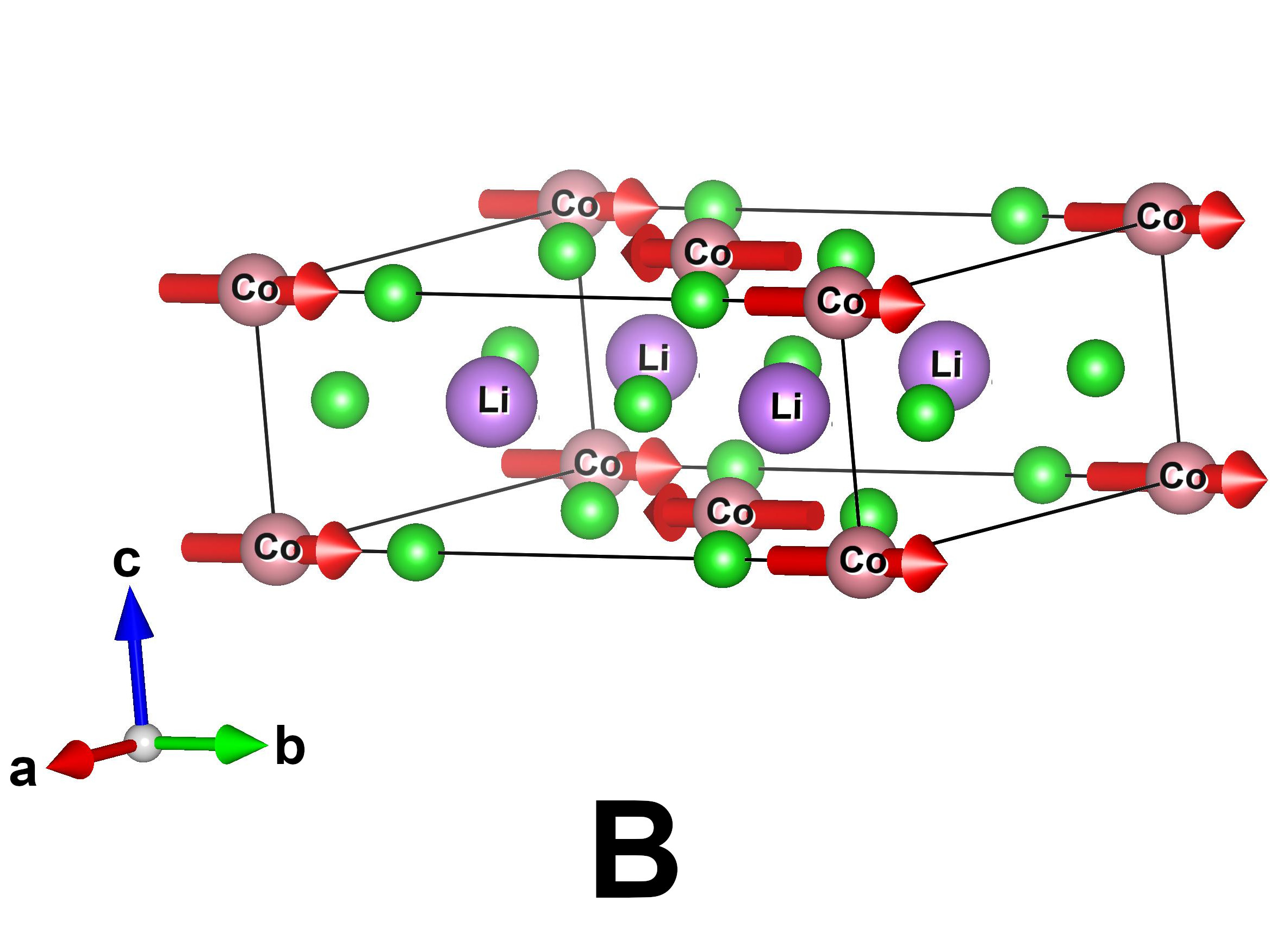}}
    \hfill
    \subfloat{\includegraphics[width=0.48\columnwidth]{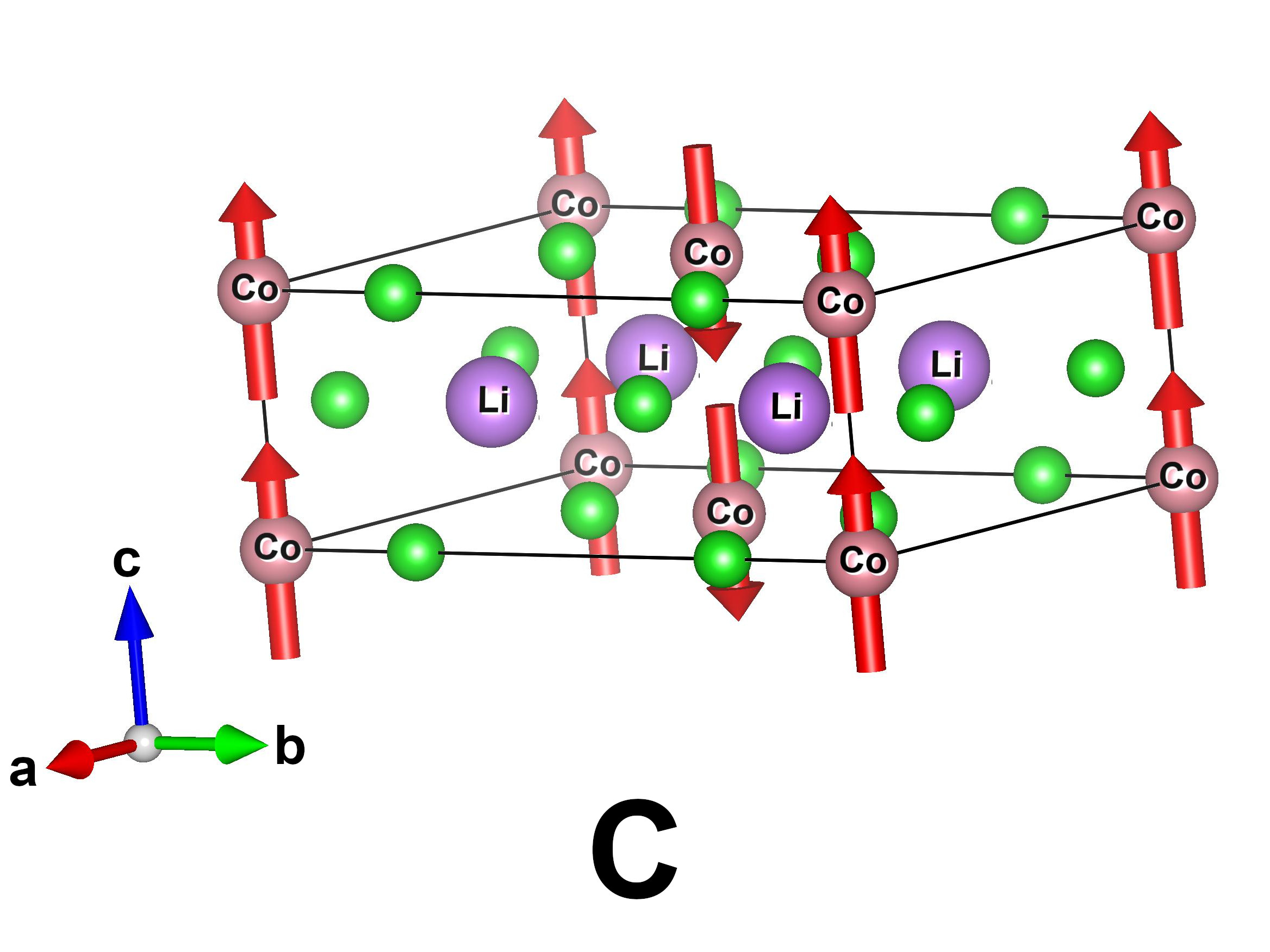}}
    \hfill
    \subfloat{\includegraphics[width=0.48\columnwidth]{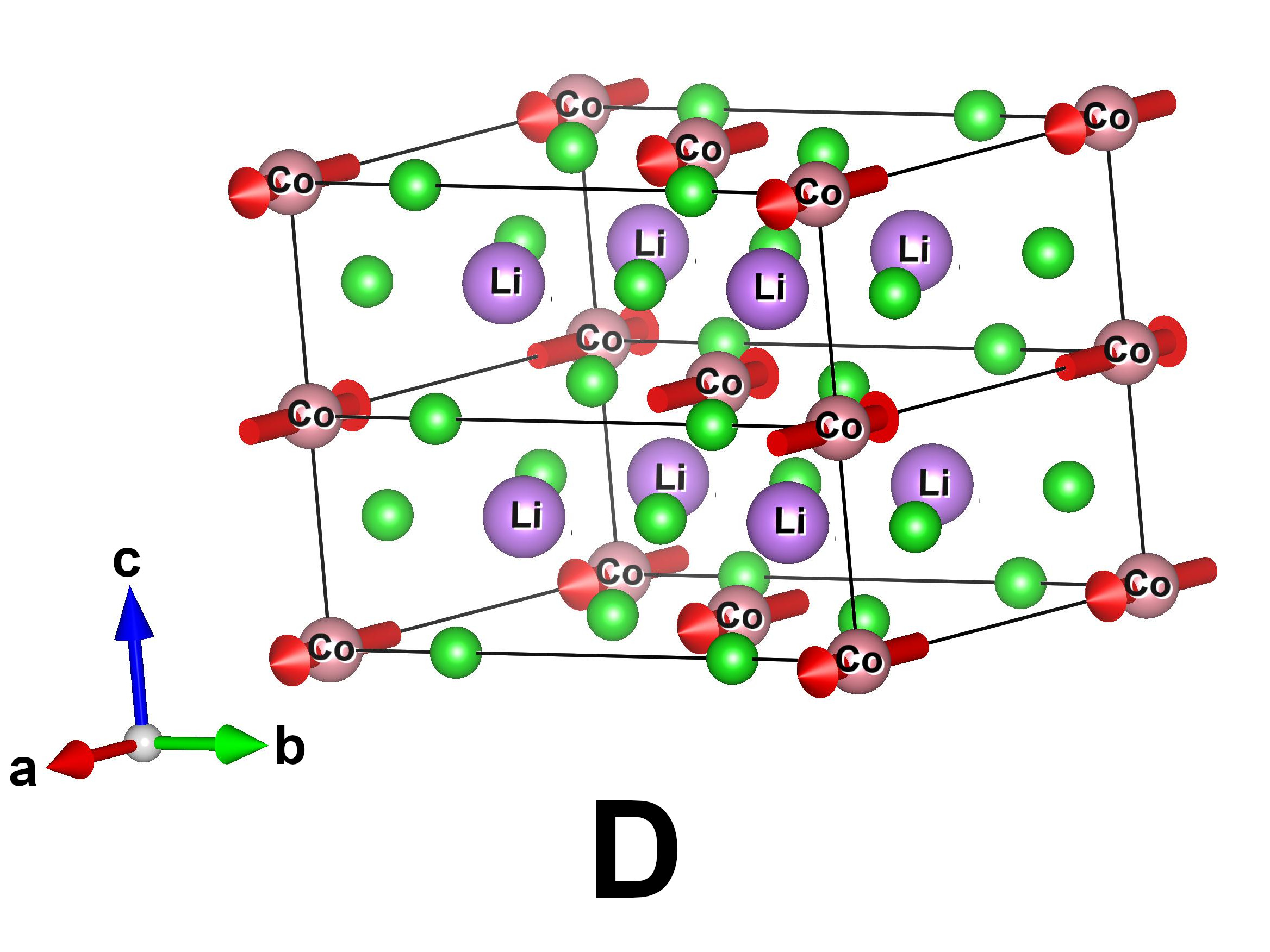}}
    \caption{The DFT-calculated highest energy configuration (D) and the two lowest energy configurations (B and C) for the relaxed \LCC\ cell are shown. B and C have ferromagnetic chains with antiferromagnetic interactions between them while D reverses this with antiferromagnetic intrachain and ferromagnetic interchain interactions. C is 0.028~meV/atom higher in energy than B, and D is 1.245~meV/atom higher in energy than B.}
    \label{fig:dftFinalCells}
\end{figure}

Other alkali-cobalt-halide materials have antiferromagnetic chains, matching the D-I configurations, while CoCl$_{2}\cdot$2H$_2$O has ferromagnetic chains with antiferromagnetic interactions between, matching the A-C configurations.\cite{narath1964antiferromagnetism} For \LCC, the B and C configurations are the lowest energy, with B only 0.028~meV/atom lower in energy than C. This adds more evidence that the \LCC\ magnetic phases match that of CoCl$_{2}\cdot$2H$_2$O. The highest and lowest energy configurations are shown in Fig.~\ref{fig:dftFinalCells}. For the 12 configurations, the calculated cobalt moment ranged from 2.442 to 2.455~$\mu_{\mathrm{B}}$. The lowest energy interaction type is the same for the unrelaxed cell as well, though the C configuration becomes the lowest energy configuration (0.062~meV/atom lower energy than B) and the hard and easy axis trends change for the D-F and G-I interaction sets.\cite{supplement}

\begin{figure}
    \centering
    \includegraphics[width=0.9\columnwidth]{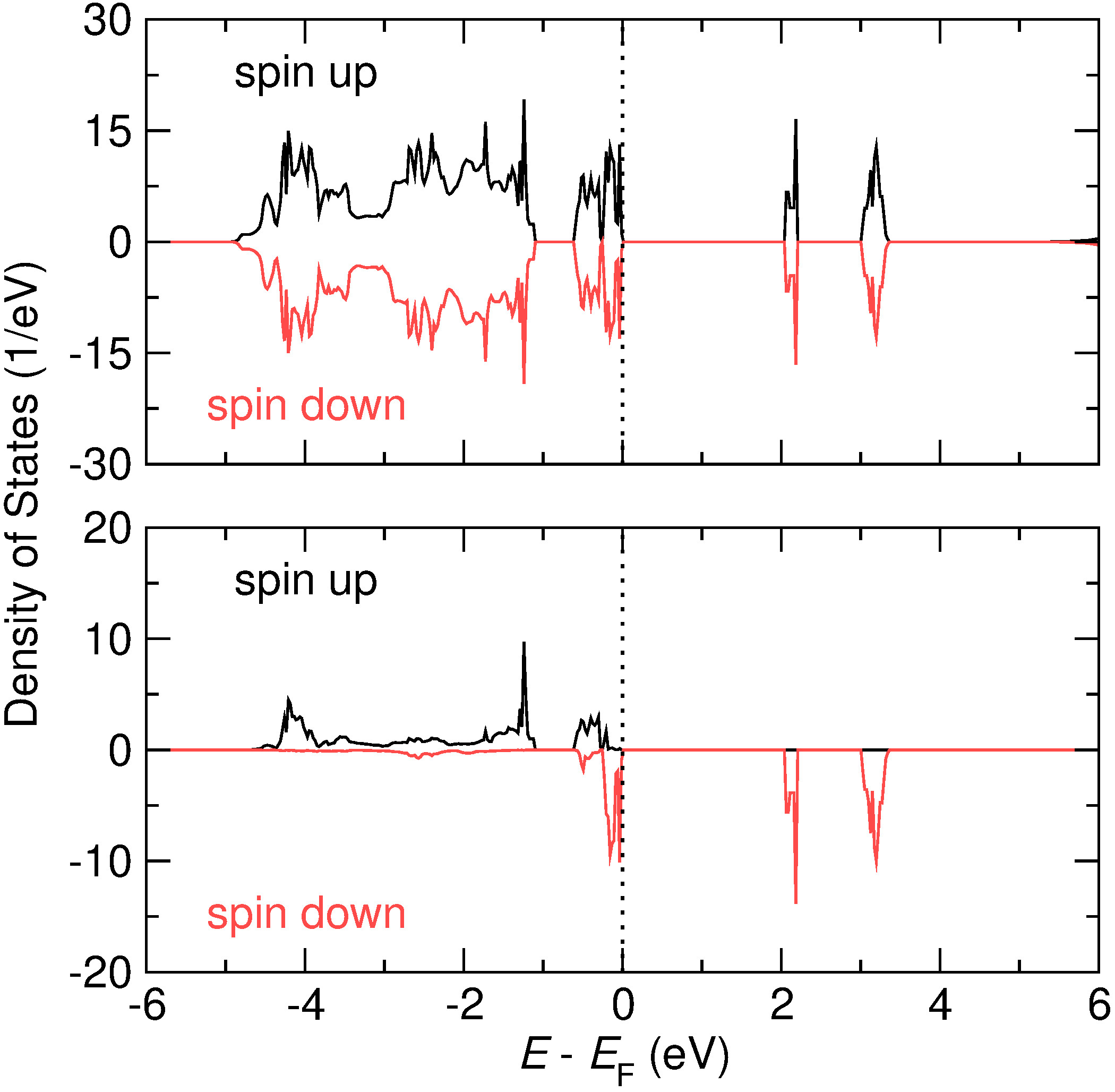}
    \caption{(Top) The total density of states for the B configuration of \LCC, calculated with the SCAN functional, is shown along with (Bottom) the density of states for a single cobalt ion's d electrons. The Fermi energy is set as the energy of the highest occupied state.}
    \label{fig:dftDOS}
\end{figure}

The density of states (DOS) for the lowest energy configuration, B, was calculated using the parameter-free SCAN functional to determine if DFT predicts \mbox{spin-$\frac{1}{2}$} or \mbox{spin-$\frac{3}{2}$} for \LCC. The total DOS and one cobalt atom's d electron projected DOS are shown in Fig.~\ref{fig:dftDOS}. Each cobalt atom's d electron projected DOS looks similar. The spin-up states are fully occupied (see bottom panel of Fig.~\ref{fig:dftDOS}), as expected for the high-spin configuration. One would also expect the spin-up minus spin-down occupation of cobalt d electrons to be 1 for \mbox{spin-$\frac{1}{2}$} and 3 for \mbox{spin-$\frac{3}{2}$} if the system were fully ionic. Integrating the states below the Fermi energy gives 4.60 spin-up electrons and 2.04 spin-down electrons. The difference, 2.56, is closer to the expected value for spin-$\frac{3}{2}$. The corresponding band structure is in the Supplemental Material.\cite{supplement}

\subsection{Zero-field magnetic structure}
While collecting zero-field neutron diffraction data, we observed several magnetic peaks after cooling from 15~K to 3.5~K. These peaks could be indexed to the nuclear unit cell parameters with a loss of \textit{C}-centering. Therefore, we used a \textbf{k}~=~(1,0,0) magnetic propagation vector to find the magnetic space group \textit{P$_C$bam} (BNS \#53.363). The neutron diffraction data refinement at 3.5~K is in Fig.~\ref{fig:ZFNPD}, and the magnetic unit cell is in Fig.~\ref{fig:magcell}. The 15~K data is in the Supplemental Material.\cite{supplement} The refined moments are parallel to the \textit{c}-axis by symmetry with ferromagnetic chains and antiferromagnetic interactions along the cell diagonal. This matches the DFT-calculated C configuration and the interaction behavior of the CoCl$_{2}\cdot$2H$_2$O low-field, antiferromagnetic phase but with a different moment axis. The interaction types further support that \LCC\ transitions from AFM to SF or FiM to PM with increasing field. As the field increases, additional ferromagnetic chains can flip throughout a powder sample, leading to the gradual transitions observed with magnetic susceptibility and magnetization measurements as well as the short-range order behavior of the magnetic heat capacity.

The refined moment is 2.19(4)~$\mu_{\mathrm{B}}$. Since neutron diffraction measures \textit{M}$_{\mathrm{sat}}$~=~\textit{gS}$\mu_{\mathrm{B}}$, we would expect a 3~$\mu_{\mathrm{B}}$ moment for \mbox{spin-$\frac{3}{2}$} assuming \textit{g}~=~2. The observed reduced moment can be explained by local defects disrupting the cobalt chains. A similar neutron diffraction refined moment has been observed in high-spin Co$^{2+}$ double-perovskite oxides with regular octahedra, which have neutron diffraction refined moments ranging from 2 to 2.43~$\mu_{\mathrm{B}}$.\cite{martinez2002preparation,viola2003structure,ortega2005factors,ivanov2010structural} The cell refinement to the neutron data shows no significant distortion of the cobalt octahedra at 3.5~K, indicating no spin-crossover from high-spin to low-spin has occurred. 

\begin{figure}
    \centering\includegraphics[width=\columnwidth]{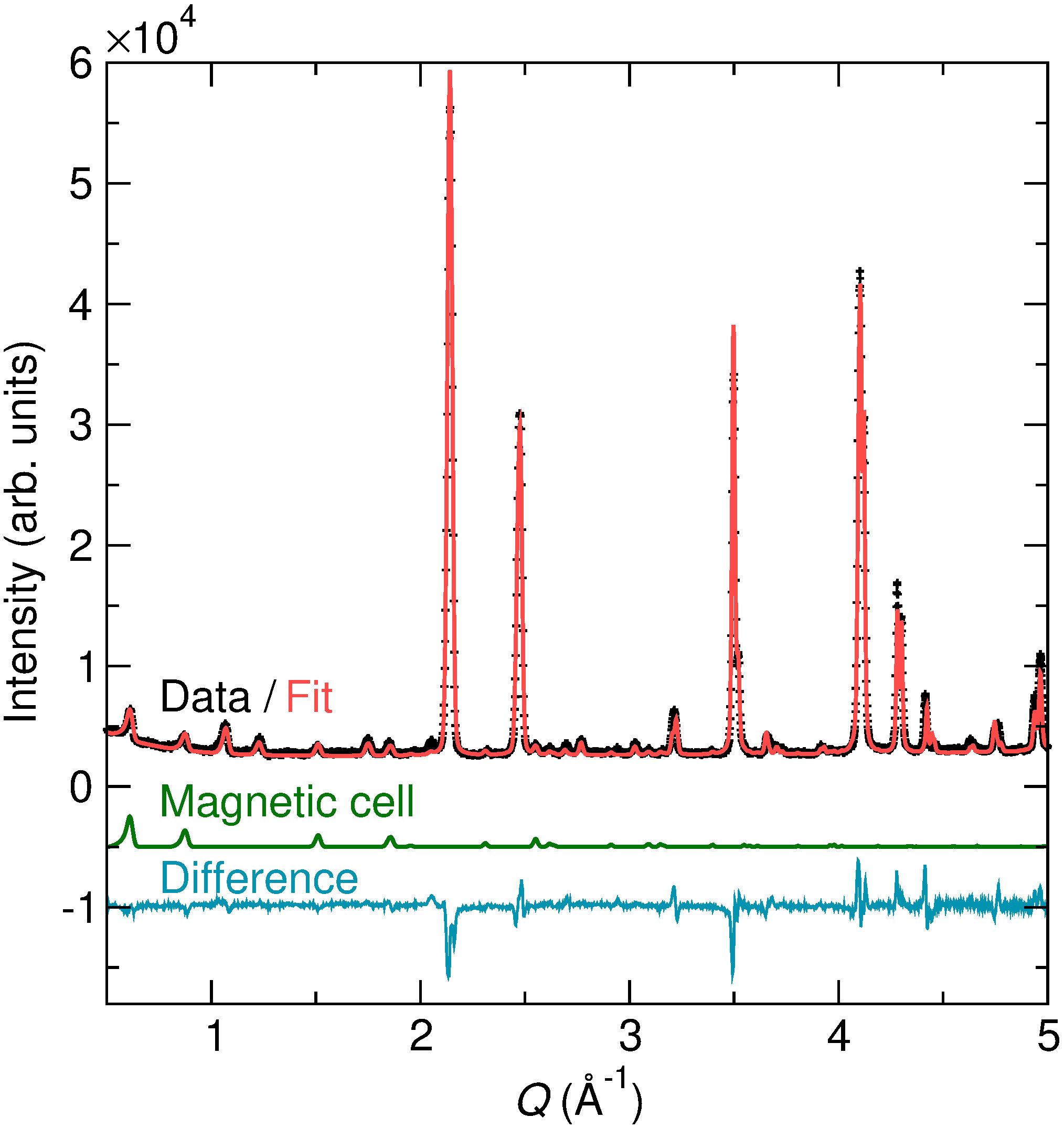}
    \caption{Zero-field neutron powder diffraction data collected on the ECHIDNA beamline at 3.5~K is refined, and the magnetic contribution is plotted.}
    \label{fig:ZFNPD}
\end{figure} 

\begin{figure}
    \centering\includegraphics[width=0.9\columnwidth]{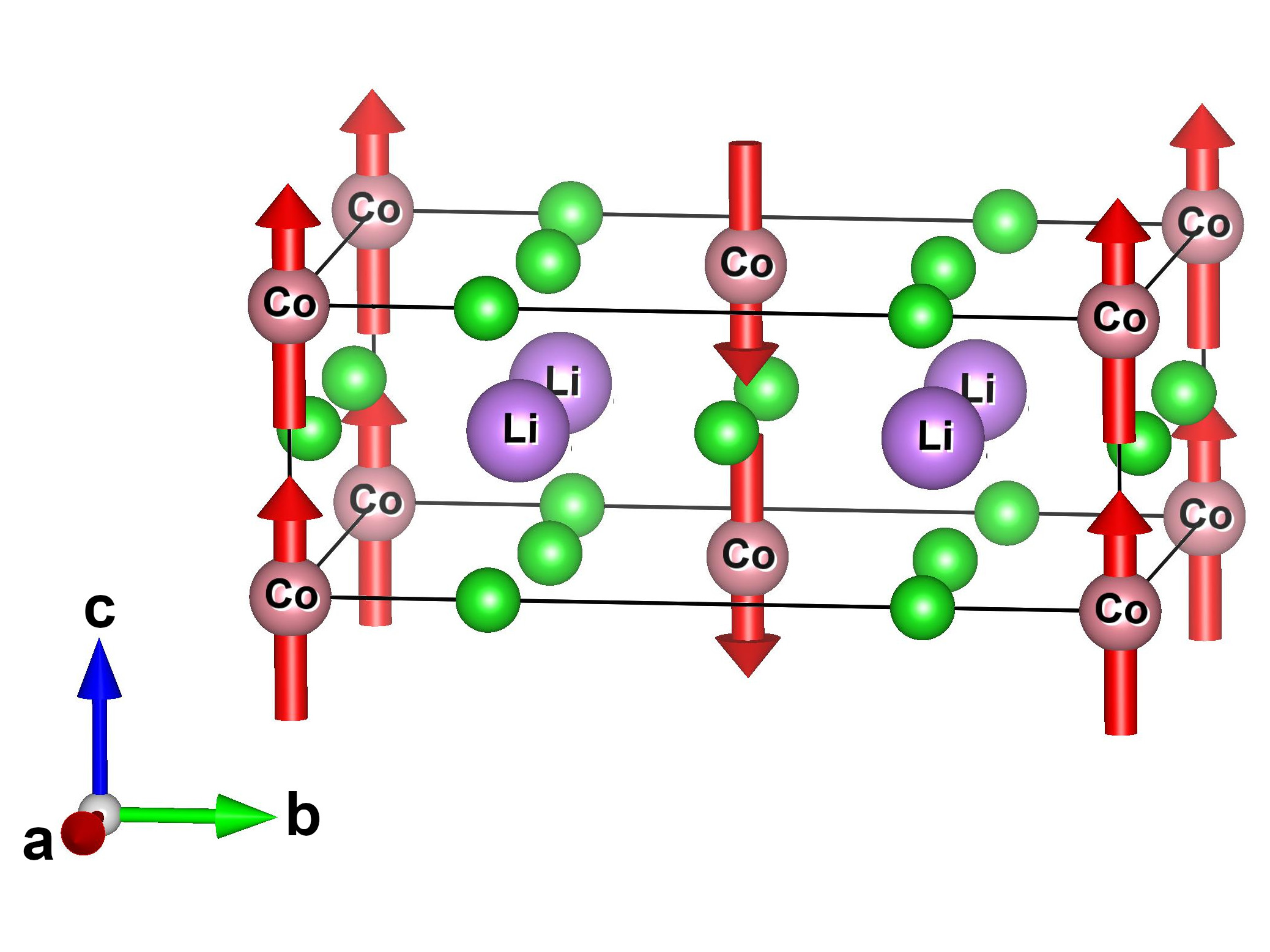}
    \caption{The zero-field magnetic unit cell of \LCC\ is commensurate with the nuclear cell with a loss of $C$-centering. Co$^{2+}$ magnetic moments with a magnitude of 2.19(4)~$\mu_{\mathrm{B}}$ are aligned along the cobalt chains.}
    \label{fig:magcell}
\end{figure} 

\subsection{Cobalt spin state}
We have presented several indicators for the spin state of Co$^{2+}$ in \LCC. High temperature data unambiguously points to a high-spin state. Curie-Weiss fits give effective moments much too large to be low-spin. DFT results, which predicted the correct interaction types, show a 2.442 to 2.455~$\mu_{\mathrm{B}}$ moment, similar to the refined moment of 2.19~$\mu_{\mathrm{B}}$ and the saturation magnetization of 2.38~$\mu_{\mathrm{B}}$, and the DFT projected DOS for the cobalt d orbitals points to the high-spin arrangement. The system, therefore, has a \mbox{spin-$\frac{3}{2}$} configuration.

It is less clear whether the system transitions to an effective \mbox{spin-$\frac{1}{2}$} state at low temperature by preferentially populating the \textit{J}=$\frac{1}{2}$ ground state doublet. The strongest evidence is the magnetic entropy change. The entropy value of 5.46~J/mol, following $\Delta$S=\textit{R}ln(2\textit{S}+1), is much closer to the \mbox{spin-$\frac{1}{2}$} case of $\Delta$S=5.76~J/mol than to the \mbox{spin-$\frac{3}{2}$} case of $\Delta$S=11.5~J/mol. Others have tried to infer an effective \mbox{spin-$\frac{1}{2}$} state from high temperature data, extracting \textit{g}, assuming it is a constant scalar quantity, from Curie-Weiss fits when orbital contributions were significant.\cite{wang2017magnetic,pramanik2019magnetic,thota2021magnetic,zhang2022magnetic} Using our 8~kOe susceptibility data gives \textit{gS}$\mu_{\mathrm{B}}$~=~3.37~$\mu_{\mathrm{B}}$ (\textit{g}~=~6.73) for \textit{S}~=~$\frac{1}{2}$ and \textit{gS}$\mu_{\mathrm{B}}$~=~4.52~$\mu_{\mathrm{B}}$ (\textit{g}~=~3.01) for \textit{S}~=~$\frac{3}{2}$. Our low temperature refined moment is, in that case, closer to the \mbox{spin-$\frac{1}{2}$} value though still $\sim$1~$\mu_{\mathrm{B}}$ different. Without spectroscopic data, neither of these effective spin-$\frac{1}{2}$ indicators is conclusive. 

\section{Conclusions}
We mapped the magnetic phase regions of \LCC\ by combining susceptibility, magnetization, and heat capacity data. The material has a low field antiferromagnetic phase, an intermediate field spin-flop or ferrimagnetic phase, and a high field paramagnetic phase. The field-driven transitions between each phase are gradual as the ferromagnetic cobalt chains rotate throughout the powder samples with increasing applied field. This leads to short range order observed as humps in the magnetic heat capacity. 

DFT calculations provided predictions for the zero-field magnetic structure within narrow energy windows. The DFT structure gave a ground state of ferromagnetic cobalt chains with antiferromagnetic interactions between them, matching the interaction types of the zero-field antiferromagnetic structure found with neutron powder diffraction. Interestingly, this behavior deviates from other alkali-cobalt-halides, which have anti-aligned intrachain moments. Further neutron diffraction studies are planned for probing the intermediate and high field magnetic structures. 

\LCC\ has a high-spin (\mbox{spin-$\frac{3}{2}$}) arrangement based on calculated and refined moments, as well as susceptibility data. But since the cobalt chain moment structure matches that of CoCl$_{2}\cdot$2H$_2$O and since its magnetic entropy change is near \textit{R}ln(2), \LCC\ may exhibit effective \mbox{spin-$\frac{1}{2}$} behavior at low temperatures, such as a quantum critical point. Using neutron scattering with a transverse field, a quantum critical point and quantized excitations have been observed in effective \mbox{spin-$\frac{1}{2}$} (electronic \mbox{spin-$\frac{3}{2}$}) Co$^{2+}$ chain materials with weakly coupled ferromagnetic intrachain interactions paired with antiferromagnetic interchain interactions.\cite{coldea2010quantum,cabrera2014excitations,thota2021magnetic,larsen2017spin} These measurements would require a single crystal of \LCC. With the magnetic phases of \LCC\ mapped, it can now be studied in more detail as a one-dimensional magnetic material.

\begin{acknowledgments}
We acknowledge the support of the Australian Centre for Neutron Scattering, ANSTO and the Australian Government through the National Collaborative Research Infrastructure Strategy, in supporting the neutron research infrastructure used in this work via ACNS proposal 13233. 

Part of this work was supported by the U.S. DOE, Office of Science, Basic Energy Sciences, Materials Sciences and Engineering Division, under contract No. DE-SC0022060. 

The authors acknowledge the use of facilities and instrumentation supported by NSF through the University of Illinois Materials Research Science and Engineering Center DMR-1720633. This work made use of the Illinois Campus Cluster, a computing resource that is operated by the Illinois Campus Cluster Program (ICCP) in conjunction with the National Center for Supercomputing Applications (NCSA) and which is supported by funds from the University of Illinois at Urbana-Champaign. Z.W.R. was supported by the National Science Foundation under Grant No. 1922758. 
\end{acknowledgments}

\bibliography{Li2CoCl4.bib}

\end{document}

\begin{center}
\Large 
\textbf{Zero-field magnetic structure and metamagnetic phase transitions of the cobalt chain compound Li$_2$CoCl$_4$}\\
\vspace{1em}
Supplemental Material\\
\vspace{1em}
\normalsize
Zachary W. Riedel, Zhihao Jiang, Maxim Avdeev, Andr\'e Schleife, and Daniel P. Shoemaker
\end{center}

\vspace{2em}

\section{\textit{In-situ} X-ray diffraction}
We used \textit{in-situ} XRD with a \LCC\ powder sample to probe the reported nuclear cell phase transition. Heated coils above and below the sample, which was sealed in a glass capillary, controlled the temperature, which was monitored with a thermocouple. Each scan spanned 2$\theta$~=~5-29.5$^{\circ}$ with 0.02049$^{\circ}$/step and 0.5~s/step ($\approx$10~min/scan). Concurrently, the temperature increased at 1$^{\circ}$C/min to 480$^{\circ}$C, where it was held 1~h before decreasing at 1$^{\circ}$C/min to room temperature.

\begin{figure}
    \centering
    \includegraphics[width=0.55\columnwidth]{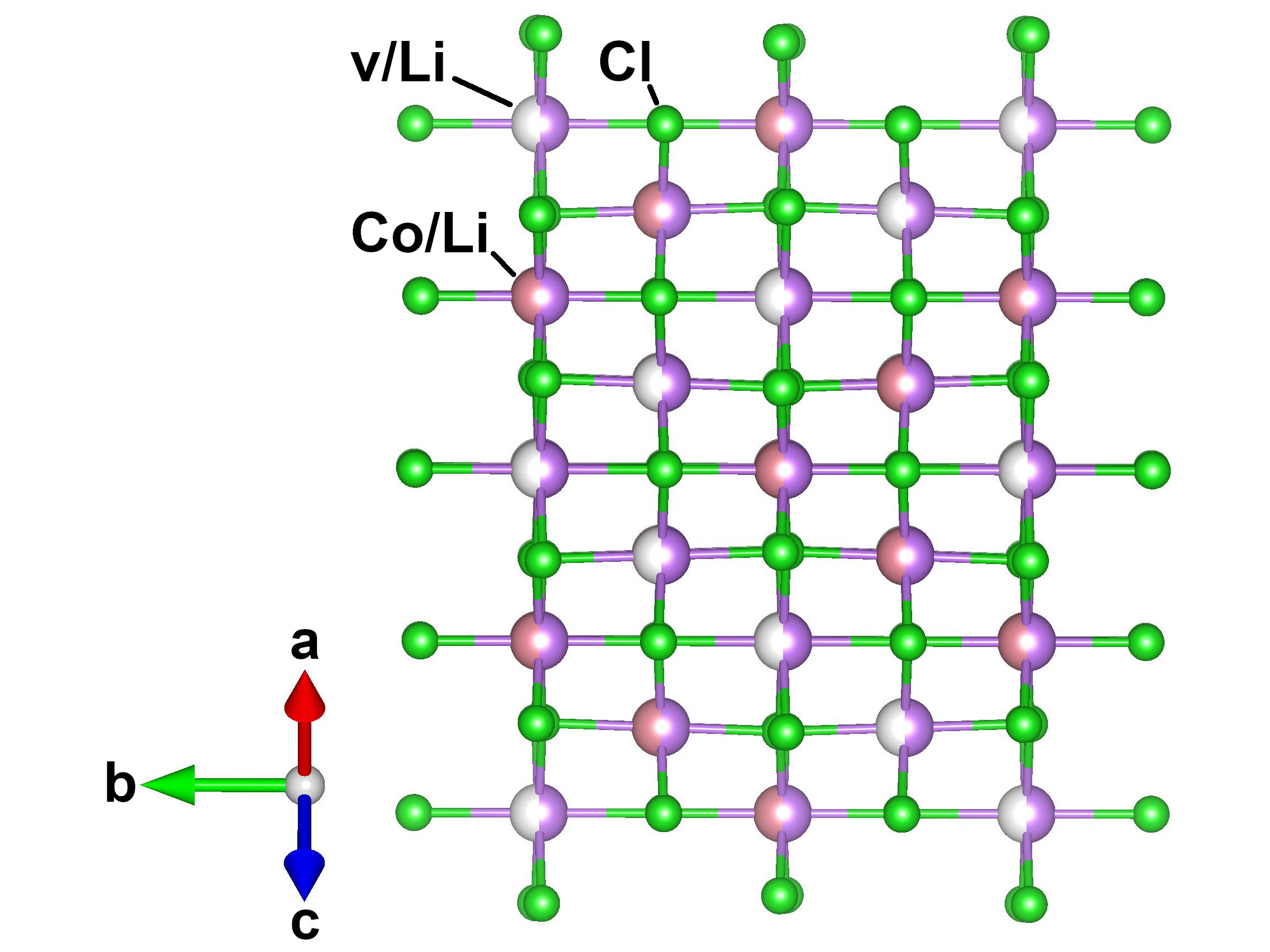}
    \caption{The high temperature cell of \LCC\ is shown. Metal site disorder fills the empty channels along the low temperature phase \textit{c}-axis.}
    \label{fig:highTcell}
\end{figure}

The high temperature phase's structure was previously determined with neutron diffraction.\cite{schneider1993polymorphie} It is a disordered rock-salt structure crystallizing in the \textit{Fd$\bar{3}$m} space group (Fig.~\ref{fig:highTcell}). Lithium occupies 1/2 of the (0,0,0) site, and lithium and cobalt evenly split occupancy of the (0.5,0.5,0.5) site. Our \textit{in-situ} XRD data show this transition near 330$^{\circ}$C during heating and cooling based on the emergence/disappearance of the low temperature (020), (200), and (001) reflections. 

\begin{figure}
    \centering
    \subfloat{\includegraphics[width=0.5\columnwidth]{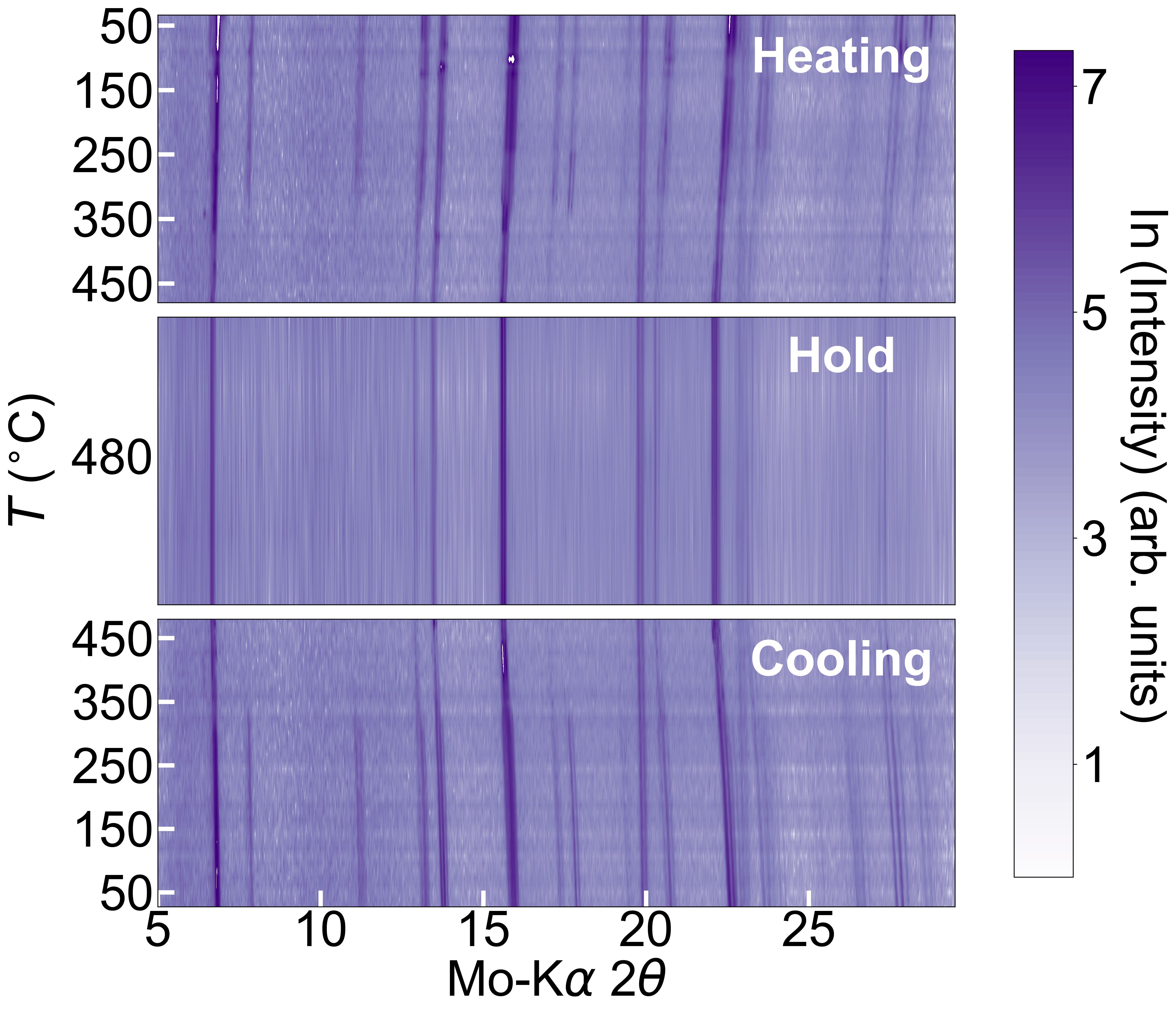}}
    \hfill
    \subfloat{\includegraphics[width=0.45\columnwidth]{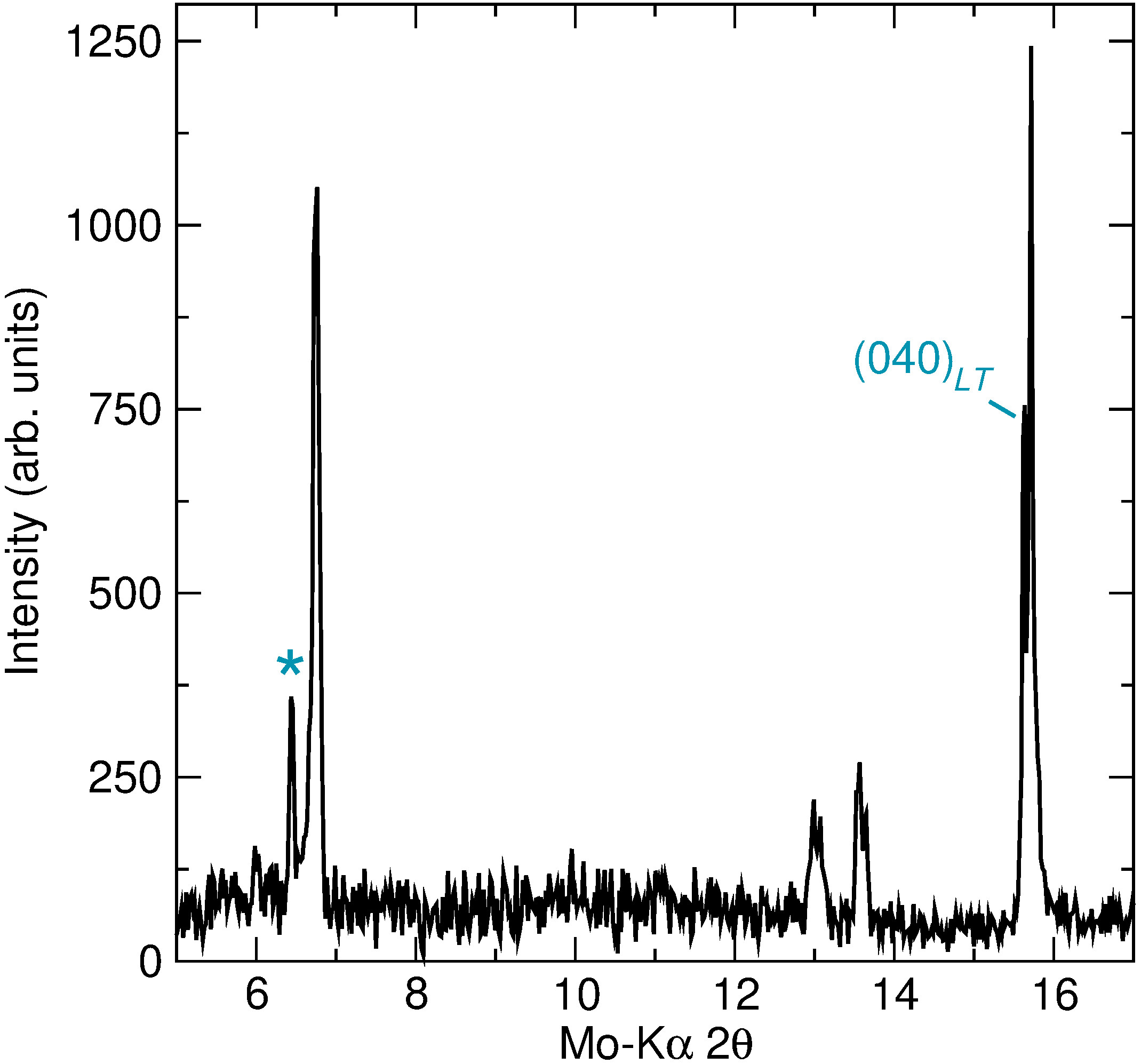}}
    \caption{(Left) \textit{In-situ} XRD upon heating, holding at 480$^{\circ}$C, and cooling is shown. (Right) The powder XRD pattern collected during the structural phase transition upon heating shows an emergent peak (*) and an increase in the (040)$_{LT}$ reflection intensity.}
    \label{fig:insitu}
\end{figure}

Two XRD pattern changes occur during the phase transition between the low temperature (LT) and high temperature (HT) phases upon heating (Fig.~\ref{fig:insitu}). First, a peak appears at \textit{d}~=~6.3~\AA. This \textit{d}-spacing is double that of the (311)$_{HT}$ reflection, which below the transition temperature splits into the (130)$_{LT}$ and (111)$_{LT}$ reflections. These reflections correspond to planes of cobalt and lithium ions as well as the centers of empty channels along the LT phase \textit{c}-axis. Second, the intensity of the (040)$_{LT}$ reflection increases before (040)$_{LT}$ and (201)$_{LT}$ merge into (400)$_{HT}$ and it drops again. Both changes may stem from the partial filling of channels that are open along the LT \textit{c}-axis since these planes correspond to distances between vacancies in the LT phase that are filled with lithium in the HT one. The changes disappear once the disordered-rock-salt structure is formed, which is within one of our ${\Delta}T{\approx}$10$^{\circ}$C scans. The additional peak was not observed upon cooling.

\pagebreak
\section{Heat capacity fit}
The difference between the constant pressure and constant volume heat capacities was estimated before fitting the constant pressure data to Debye and Einstein terms. The lattice volumes came from room temperature powder X-ray diffraction and 3.5~K powder neutron diffraction, and the density came from room temperature X-ray diffraction. The bulk modulus of 25~GPa came from the Materials Project database.\cite{jain2013commentary,de2015charting}

\begin{equation}
    C_\mathrm{p}-C_\mathrm{v} = \alpha^2 B V_0 T
\end{equation}
\begin{equation}
    \alpha_V = \frac{\Delta V}{\Delta T} \times \frac{1}{V_0} \approx \frac{(266.855-261.412) ~\mathrm{\text{\AA}^3}}{(298-3.5)~\mathrm{K}} \times \frac{1}{266.855~\mathrm{\text{\AA}^3}} = 6.9{\times}10^{-5}/\mathrm{K}
\end{equation}
\begin{equation}
    V_0 = \frac{MW}{\rho} = \frac{214.63~\mathrm{g/mol}}{2.671~\mathrm{g/cm^3}} = 80.4~\mathrm{cm^3/mol}
\end{equation}
\begin{equation}
    \textit{C}_\mathrm{p}-\textit{C}_\mathrm{v} (3.5~\mathrm{K}) \approx 0.03~\mathrm{J/mol/K}
\end{equation}
\begin{equation}
    \textit{C}_\mathrm{p}-\textit{C}_\mathrm{v} (145~\mathrm{K}) \approx 1.4~\mathrm{J/mol/K}
\end{equation}

The approximations are only 2\% of the 3.5~K constant pressure heat capacity and 0.6\% of the 145~K heat capacity at zero field and were, consequently, ignored. The best fit for the constant pressure heat capacity at zero field was found using one Debye term and two Einstein terms (Eq.~\ref{eq:Debye_Einstein}). The resulting parameters are in Table~\ref{tab:DE_fit}, and the plotted function is in the main text.
\begin{equation}
    \textit{C}_\mathrm{p} = 9{\alpha}R(\frac{T}{{\theta}_{D}})^3 \int_{0}^{{\theta}_{D}/T} \frac{x^4 e^x}{(e^x - 1)^2} dx + \sum_{j=1}^{2} 3{\beta}_{j}R(\frac{{\theta}_{Ej}}{T})^2 \frac{e^{{\theta}_{Ej}/T}}{(e^{{\theta}_{Ej}/T} - 1)^2}
    \label{eq:Debye_Einstein}
\end{equation}

\begin{table}
\small
\centering 
\caption{\label{tab:DE_fit} The fit parameters for Eq.~\ref{eq:Debye_Einstein} are shown for zero-field constant pressure heat capacity data above 15.5~K (R$^2$=0.99998).} 
\begin{tabular}{c c c}
\hline Parameter & Value & Estimated Error \\
\hline
${\theta}_D$ & 504~K & 11~K \\
${\theta}_{E1}$ & 68~K & 4~K \\
${\theta}_{E2}$ & 189~K & 8~K \\
$\alpha$ & 8.3 & 0.3 \\
${\beta}_1$ & 1.2 & 0.1 \\
${\beta}_2$ & 3.6 & 0.2 \\
\end{tabular}
\end{table}

\pagebreak
\section{Additional susceptibility and magnetization data}
\begin{figure}[h]
    \centering
    \subfloat{\includegraphics[width=0.45\columnwidth]{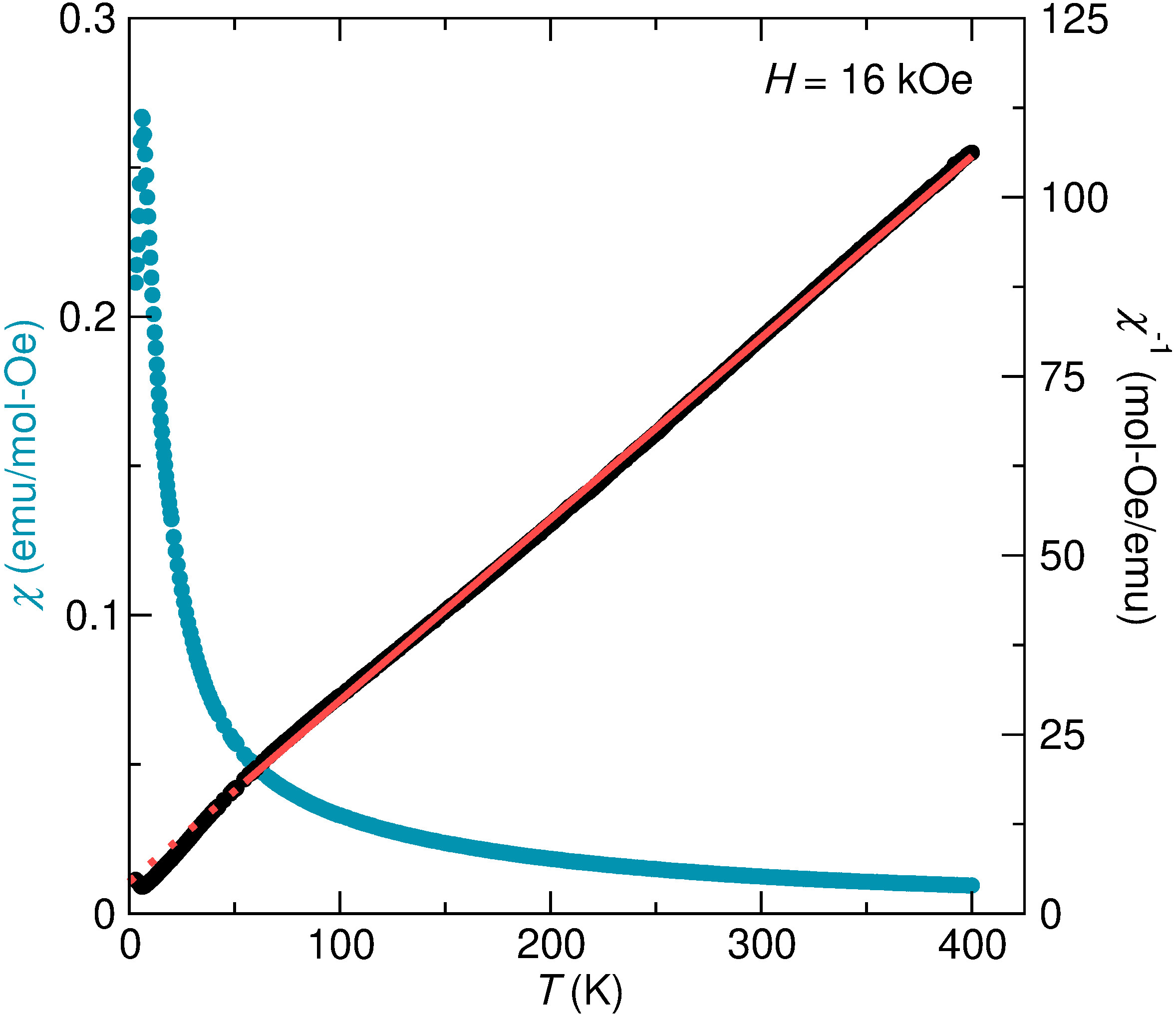}}
    \hfill
    \subfloat{\includegraphics[width=0.45\columnwidth]{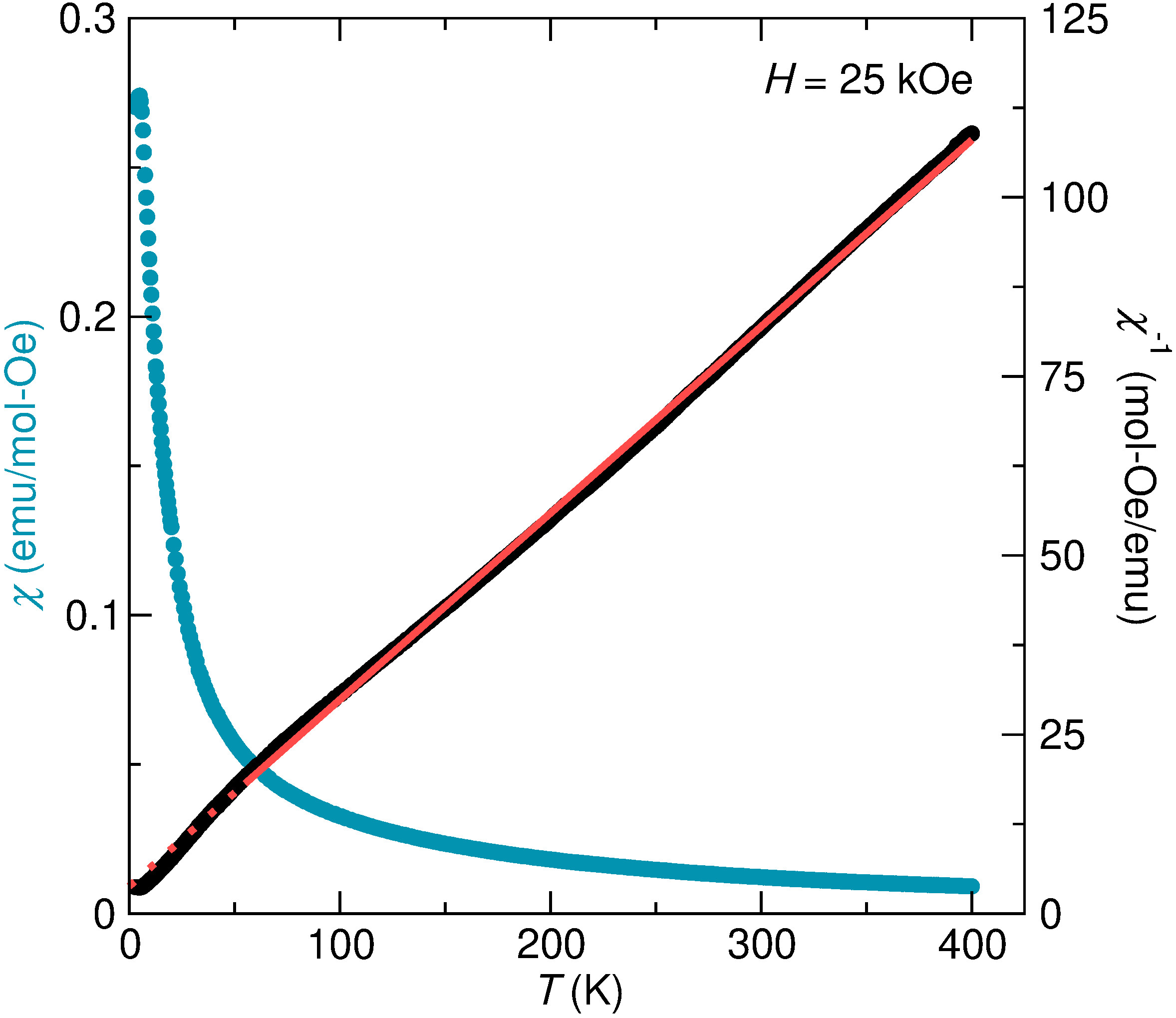}}
    \vfill
    \subfloat{\includegraphics[width=0.45\columnwidth]{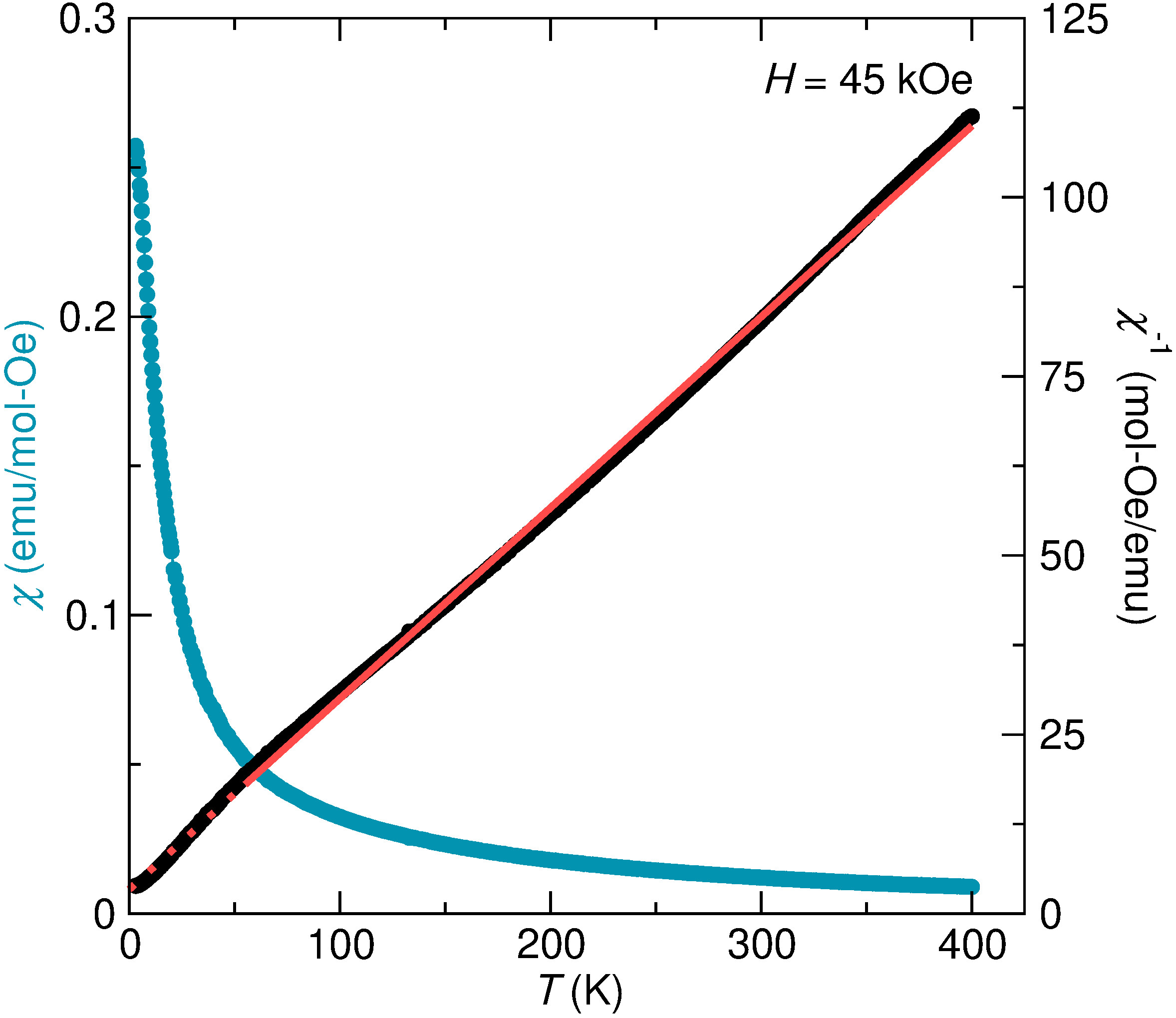}}
    \caption{The susceptibility at 16~kOe is near the low field to intermediate field transition. The 25~kOe susceptibility is in the intermediate field range. And the 45~kOe susceptibility is in the high field region.}
    \label{fig:MvTother}
\end{figure}

The field-cooled susceptibility and Curie-Weiss fits for several applied fields are shown in Fig.~\ref{fig:MvTother}. Fitting to Curie-Weiss paramagnetism from 55 to 400~K gives $\theta_{\mathrm{CW}}$~=~\mbox{-17.6}~K and $\mu_{\mathrm{eff}}$~=~5.62~$\mu_{\mathrm{B}}$ for 16~kOe, $\theta_{\mathrm{CW}}$~=~\mbox{-14.8}~K and $\mu_{\mathrm{eff}}$~=~5.54~$\mu_{\mathrm{B}}$ for 25~kOe, and $\theta_{\mathrm{CW}}$~=~\mbox{-12.8}~K and $\mu_{\mathrm{eff}}$~=~5.48~$\mu_{\mathrm{B}}$ for 45~kOe. The increasing Curie-Weiss parameter indicates that antiferromagnetic interactions steadily lose strength with increasing field, and the effective moment remains much higher than the spin-only value of 3.87~$\mu_{\mathrm{B}}$, so there remains a strong orbital contribution to the \mbox{spin-$\frac{3}{2}$} state. However, significant deviations from linear behavior that were not present in the 8~kOe data are apparent with increasing applied field. These deviations are additional evidence of ferromagnetic interactions competing with antiferromagnetic ones in the intermediate and high field regions as ferromagnetic cobalt chain moments rotate.

\begin{figure}[h]
    \centering
    \subfloat{\includegraphics[width=0.48\columnwidth]{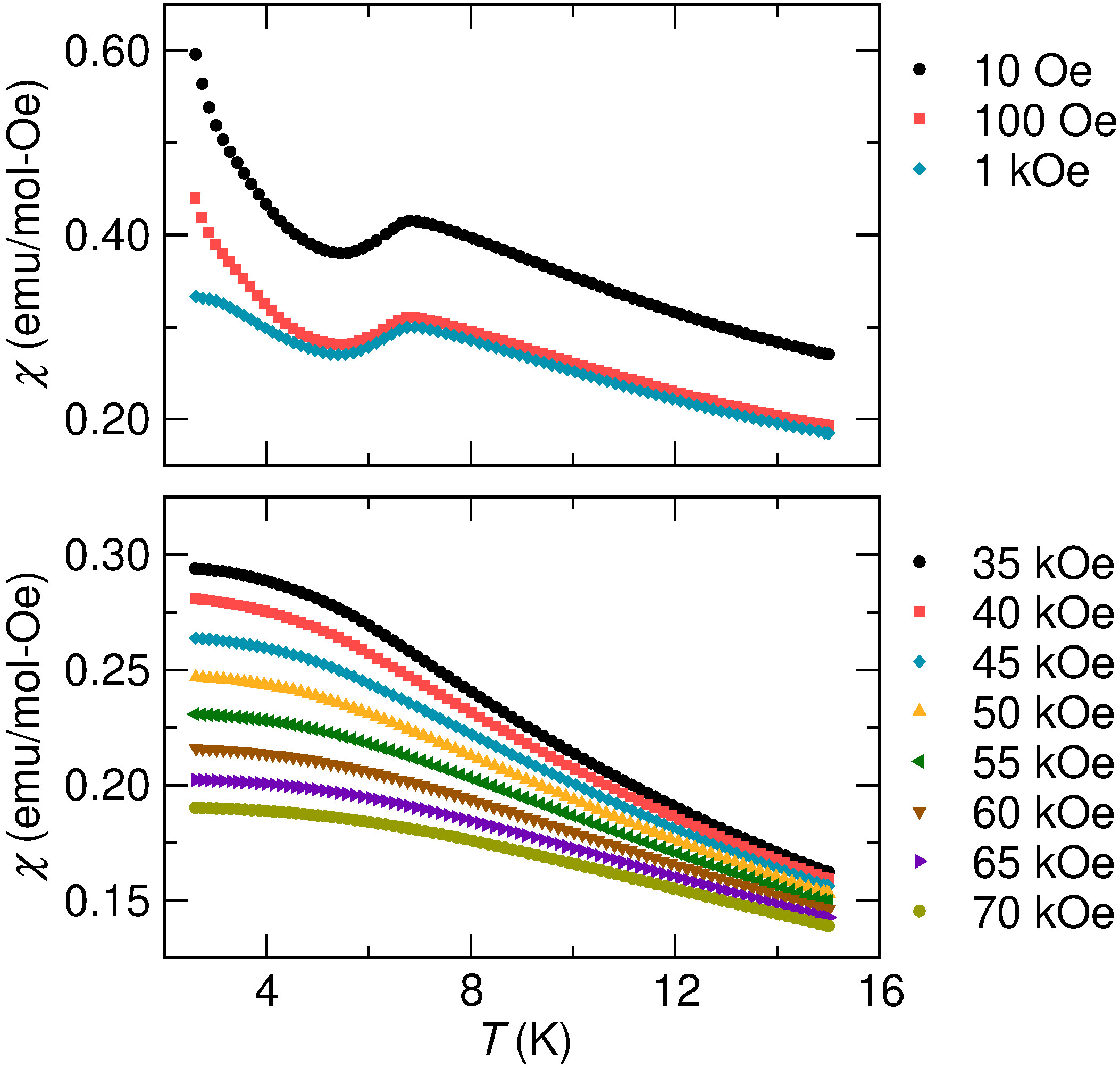}}
    \hfill
    \subfloat{\includegraphics[width=0.47\columnwidth]{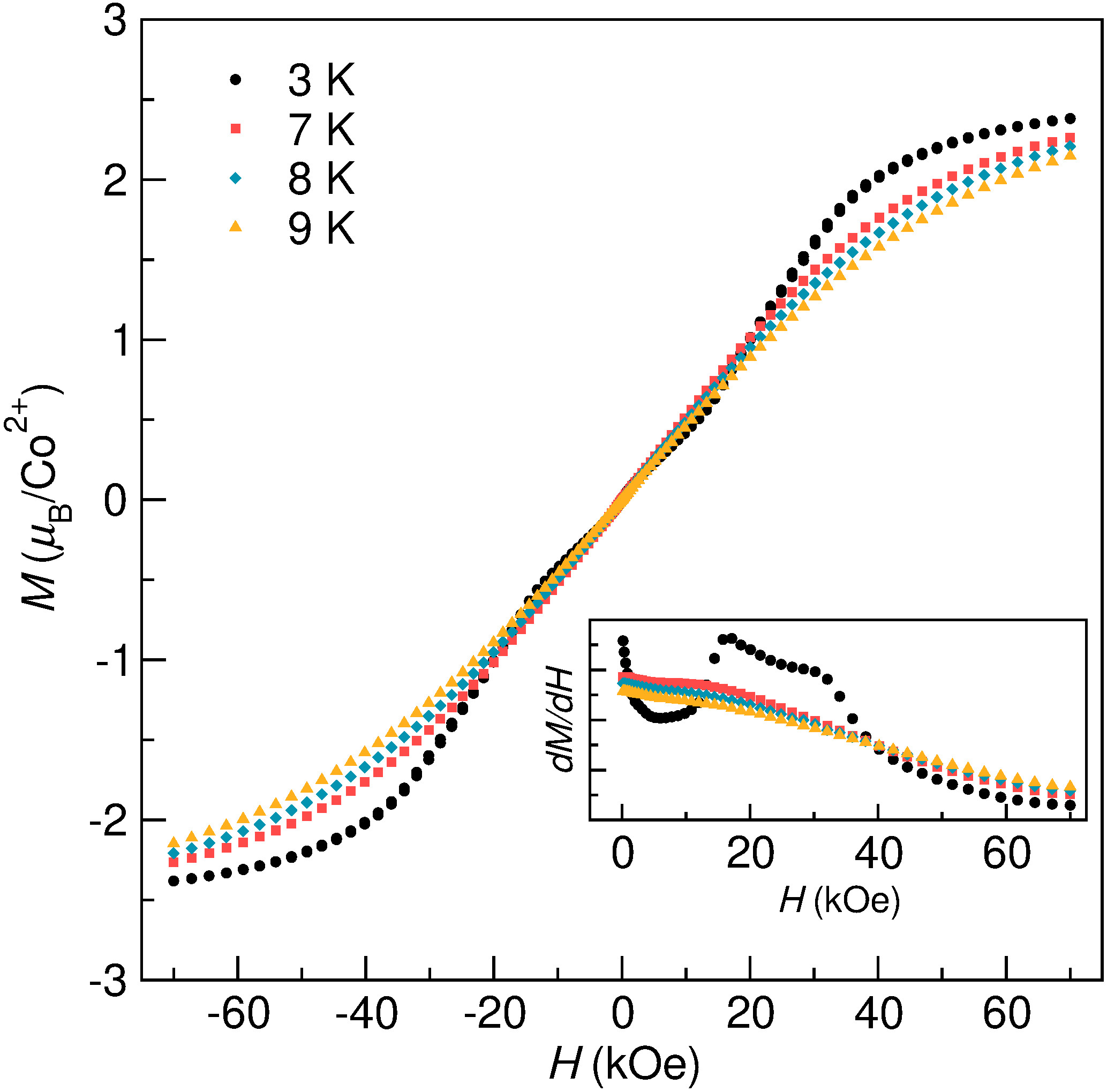}}
    \caption{Zero-field-cooled susceptibility is shown for (Top Left) low field data and for (Bottom Left) high field data. Increasing magnetic field suppresses the susceptibility. At \textit{H}$\leq$1~kOe, a tail is present at temperatures below the ordering transition. (Right) Magnetization data not presented in the main text, along with the derivative curves for the initial field increase, are shown.}
    \label{fig:ZFCother}
\end{figure}

\clearpage
The zero-field-cooled susceptibility and magnetization curves used to construct the magnetic phase diagram that are not presented in the main text are shown in Fig.~\ref{fig:ZFCother}.

\clearpage
\section{DFT magnetic moment configurations}
\begin{figure}[ht]
    \centering
    \subfloat{\includegraphics[width=0.31\columnwidth]{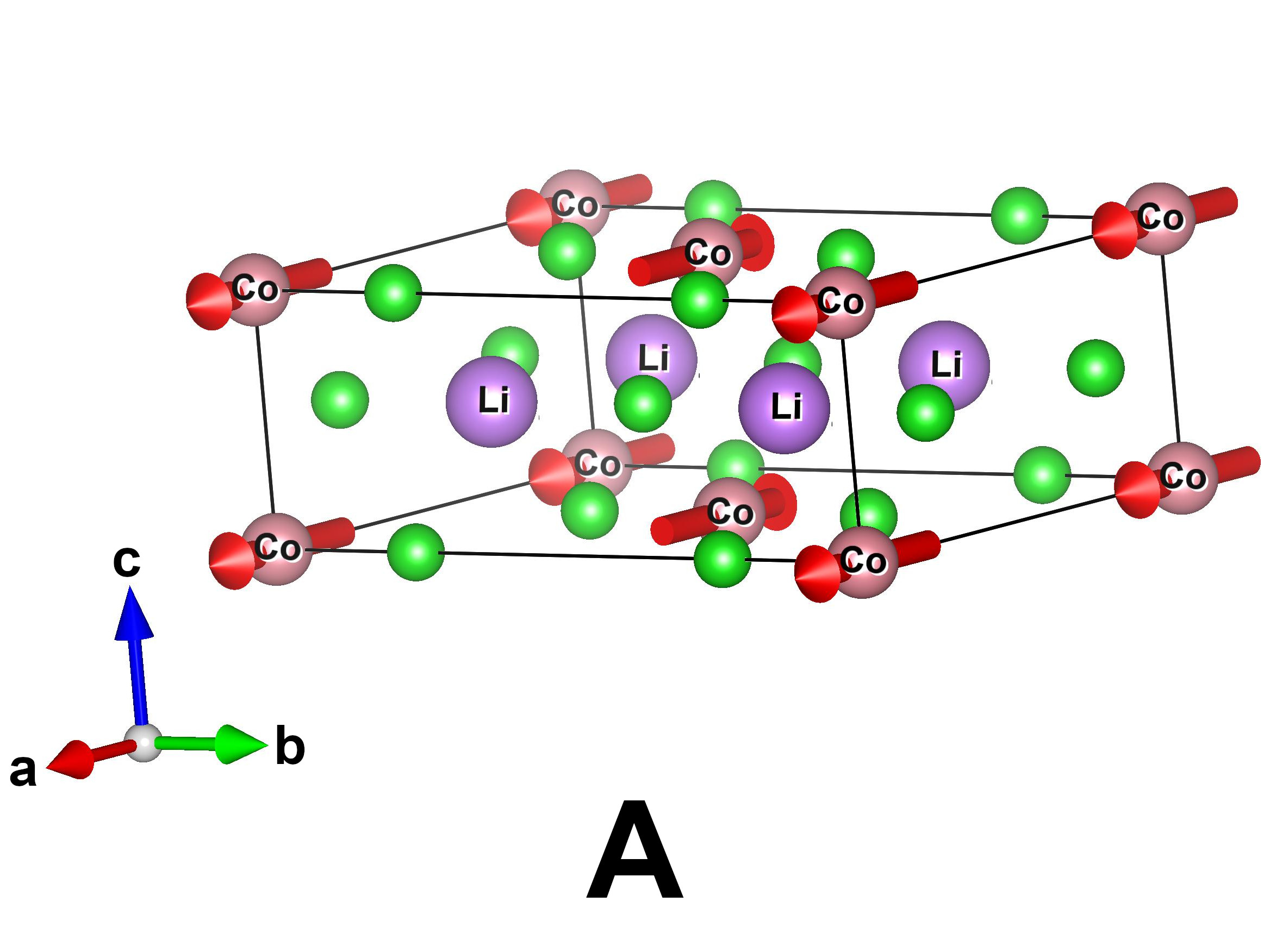}}
    \hfill
    \subfloat{\includegraphics[width=0.31\columnwidth]{Figures/Structures/dft/dft_cell_B.jpg}}
    \hfill
    \subfloat{\includegraphics[width=0.31\columnwidth]{Figures/Structures/dft/dft_cell_C.jpg}}
    \hfill
    \vspace{0.5em}
    \subfloat{\includegraphics[width=0.31\columnwidth]{Figures/Structures/dft/dft_cell_D.jpg}}
    \hfill
    \subfloat{\includegraphics[width=0.31\columnwidth]{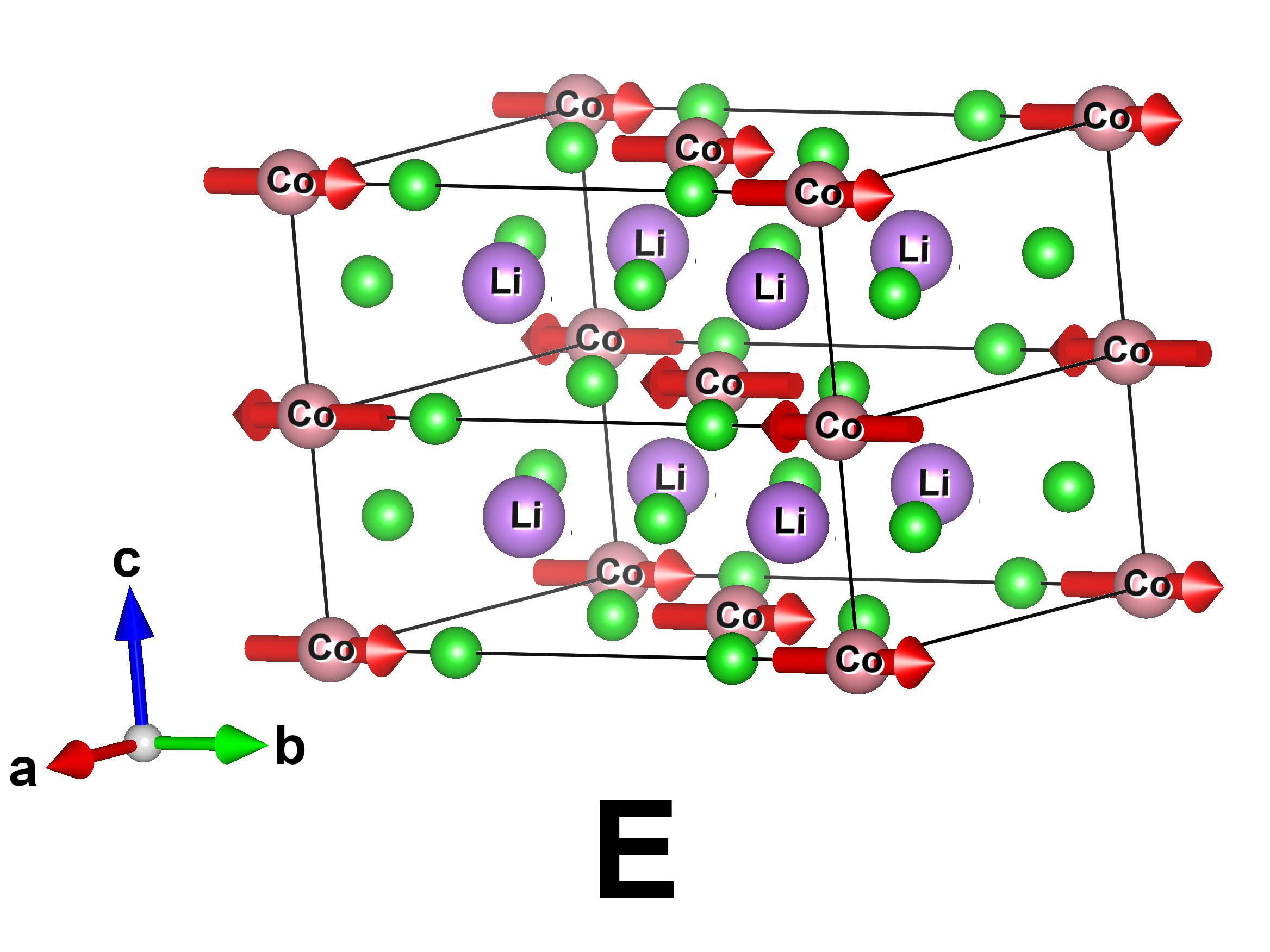}}
    \hfill
    \subfloat{\includegraphics[width=0.31\columnwidth]{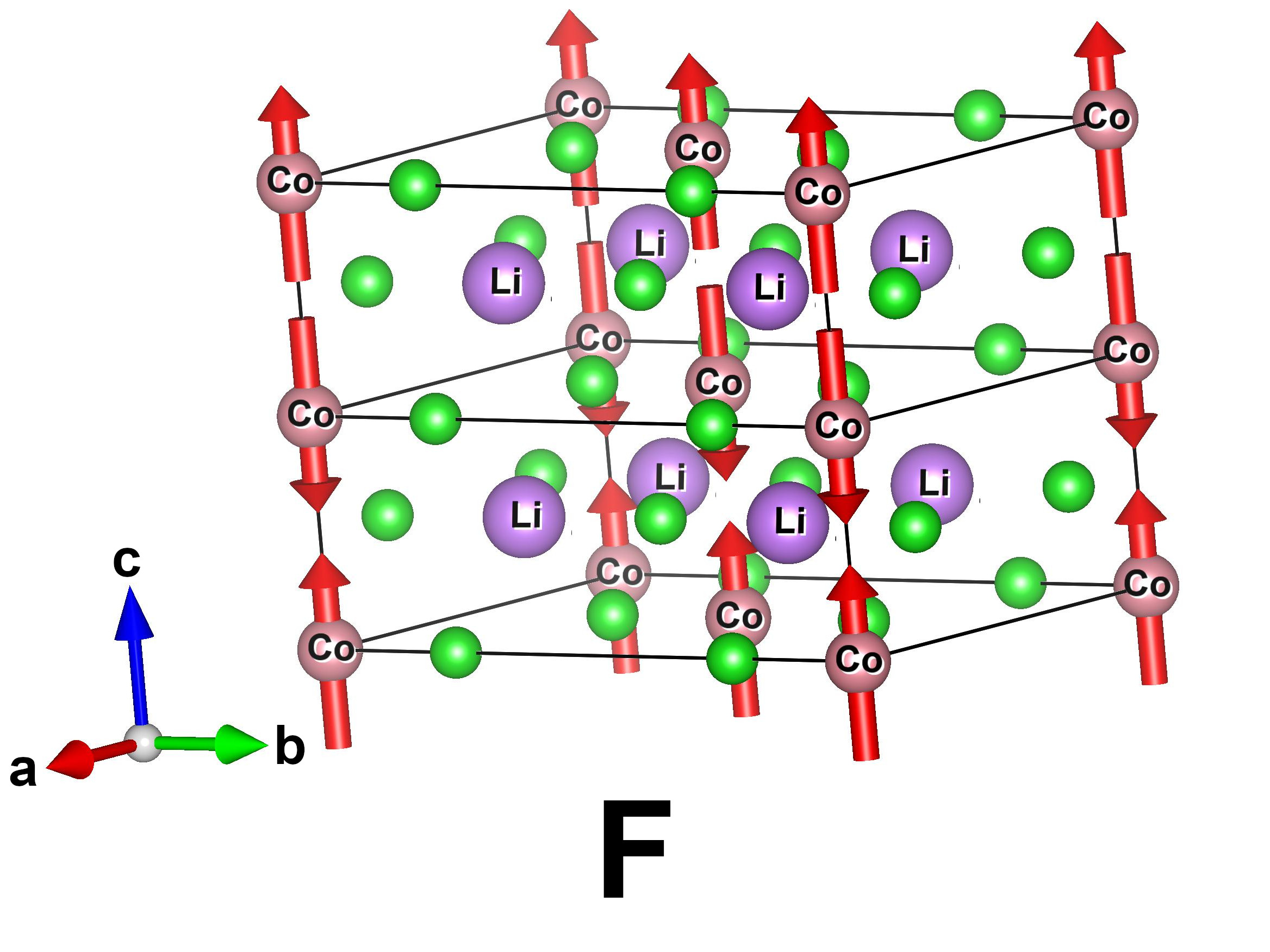}}
    \hfill
    \vspace{0.5em}
    \subfloat{\includegraphics[width=0.31\columnwidth]{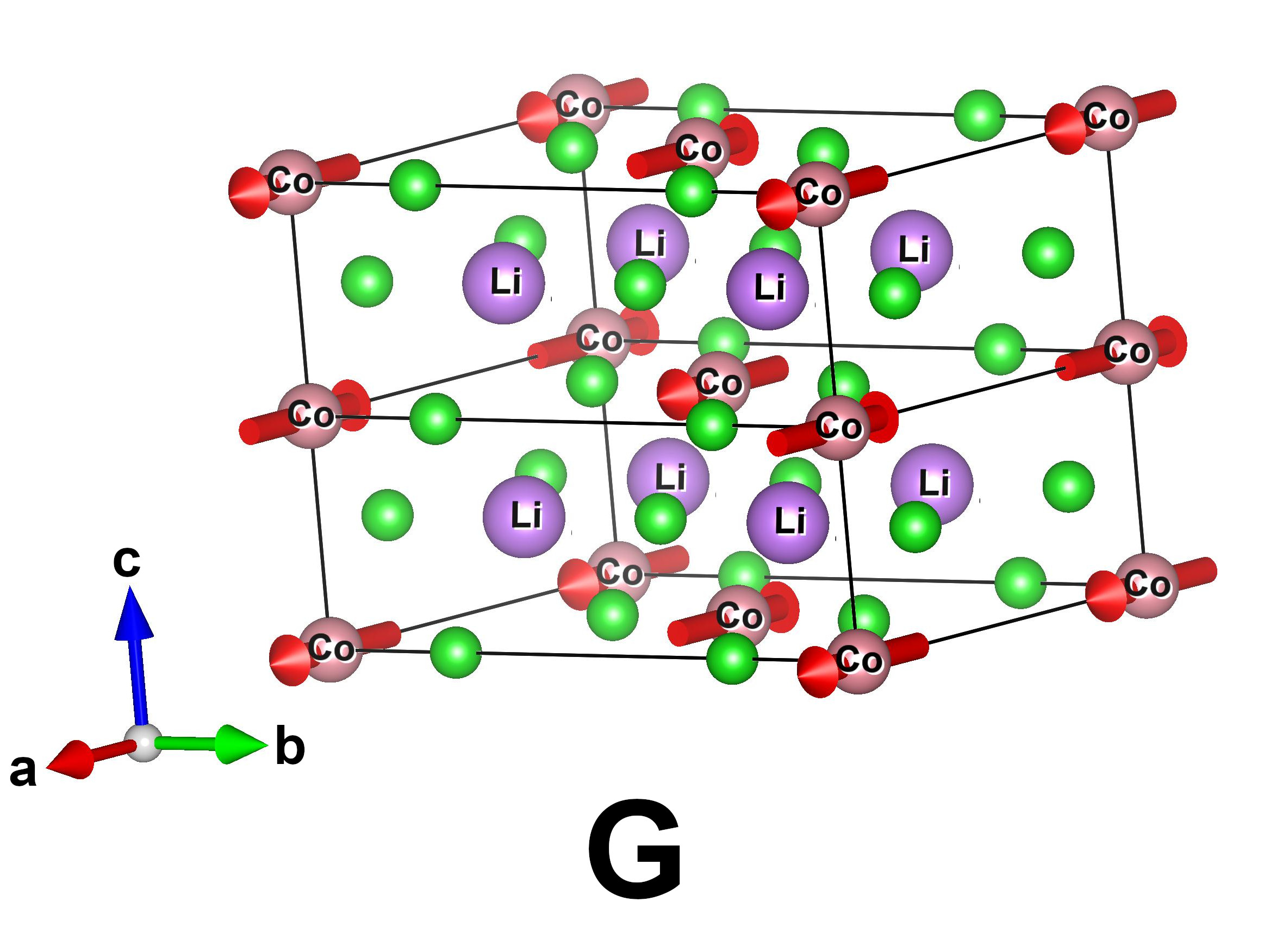}}
    \hfill
    \subfloat{\includegraphics[width=0.31\columnwidth]{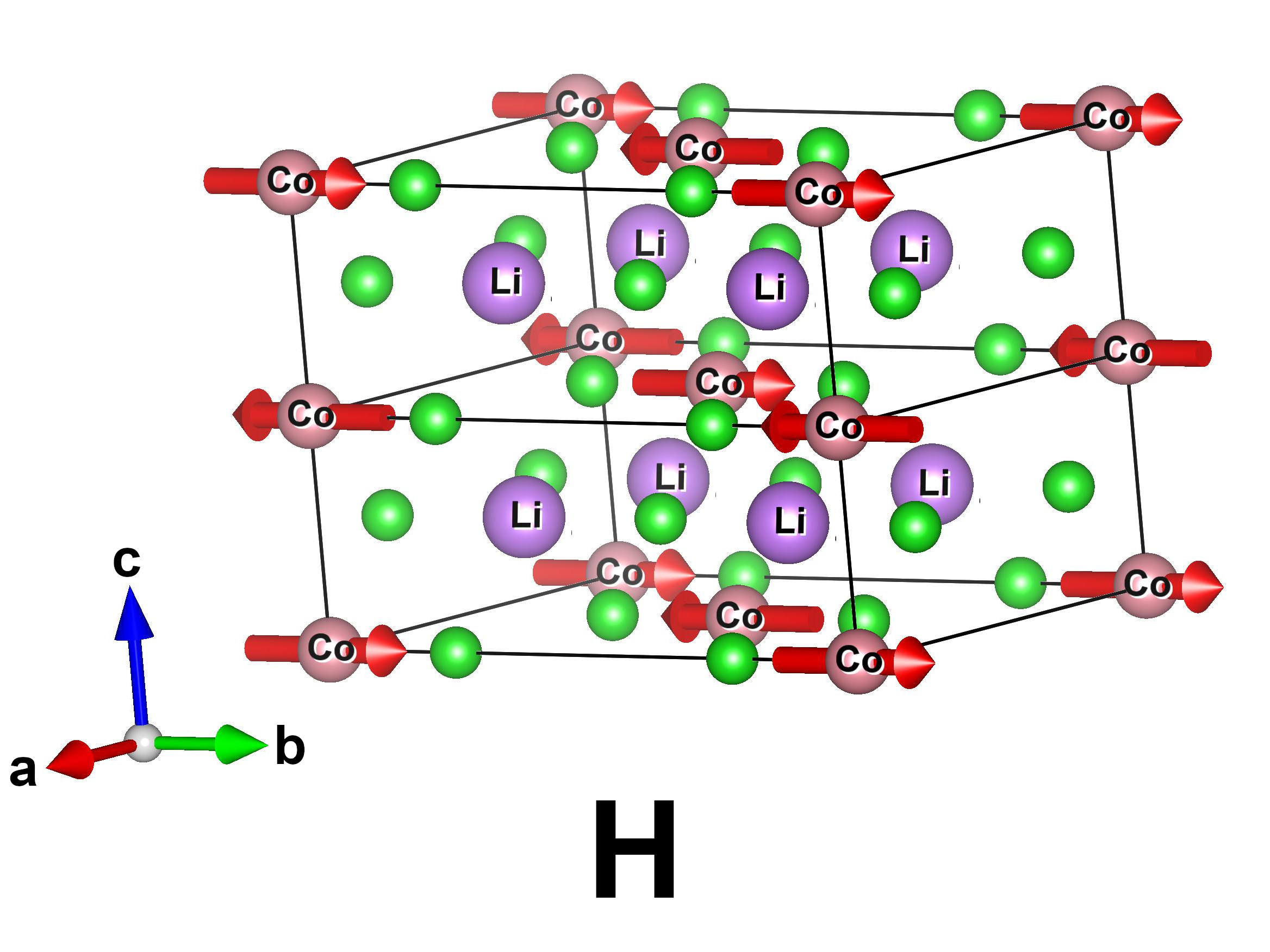}}
    \hfill
    \subfloat{\includegraphics[width=0.31\columnwidth]{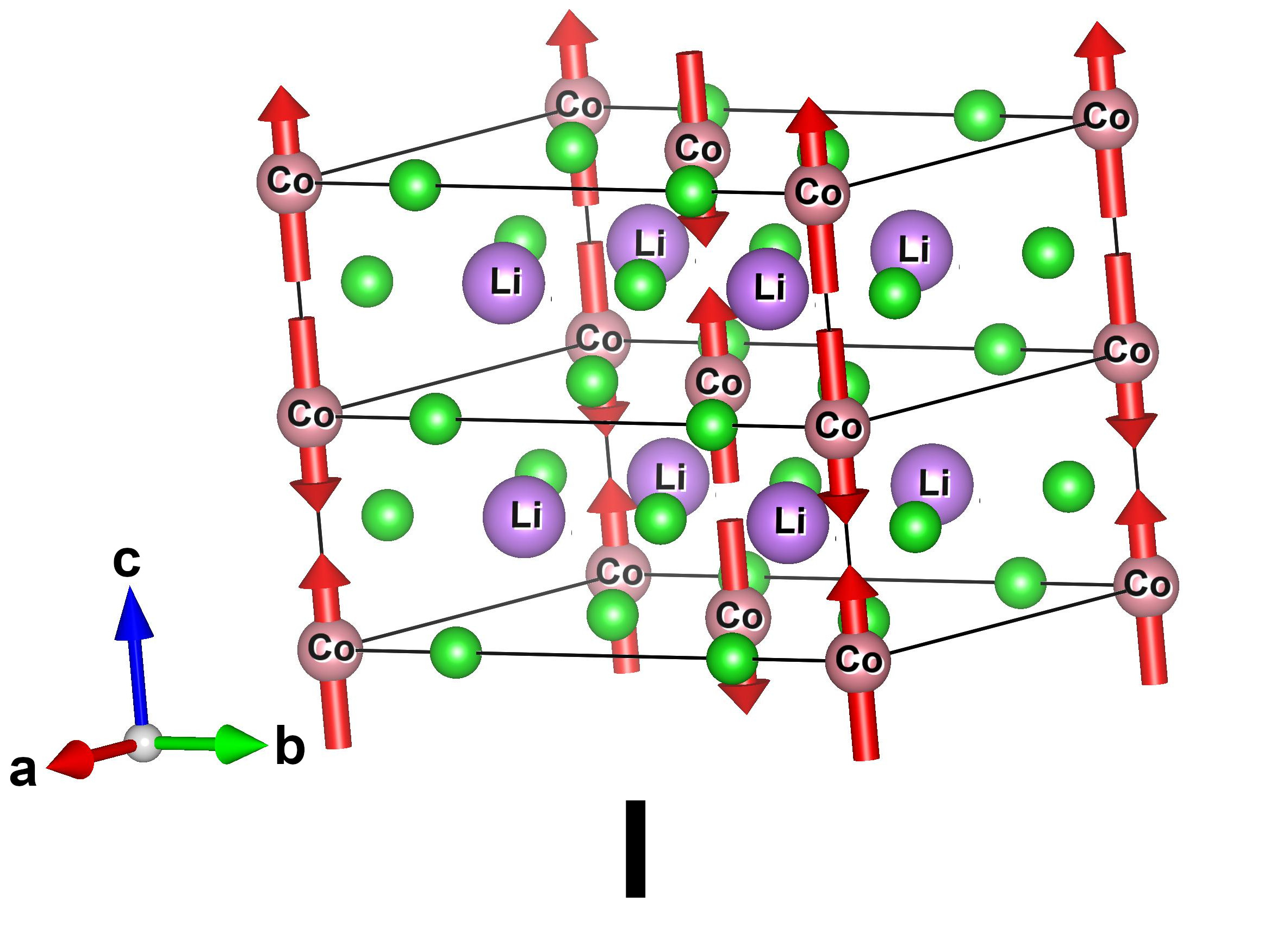}}
    \hfill
    \vspace{0.5em}
    \subfloat{\includegraphics[width=0.31\columnwidth]{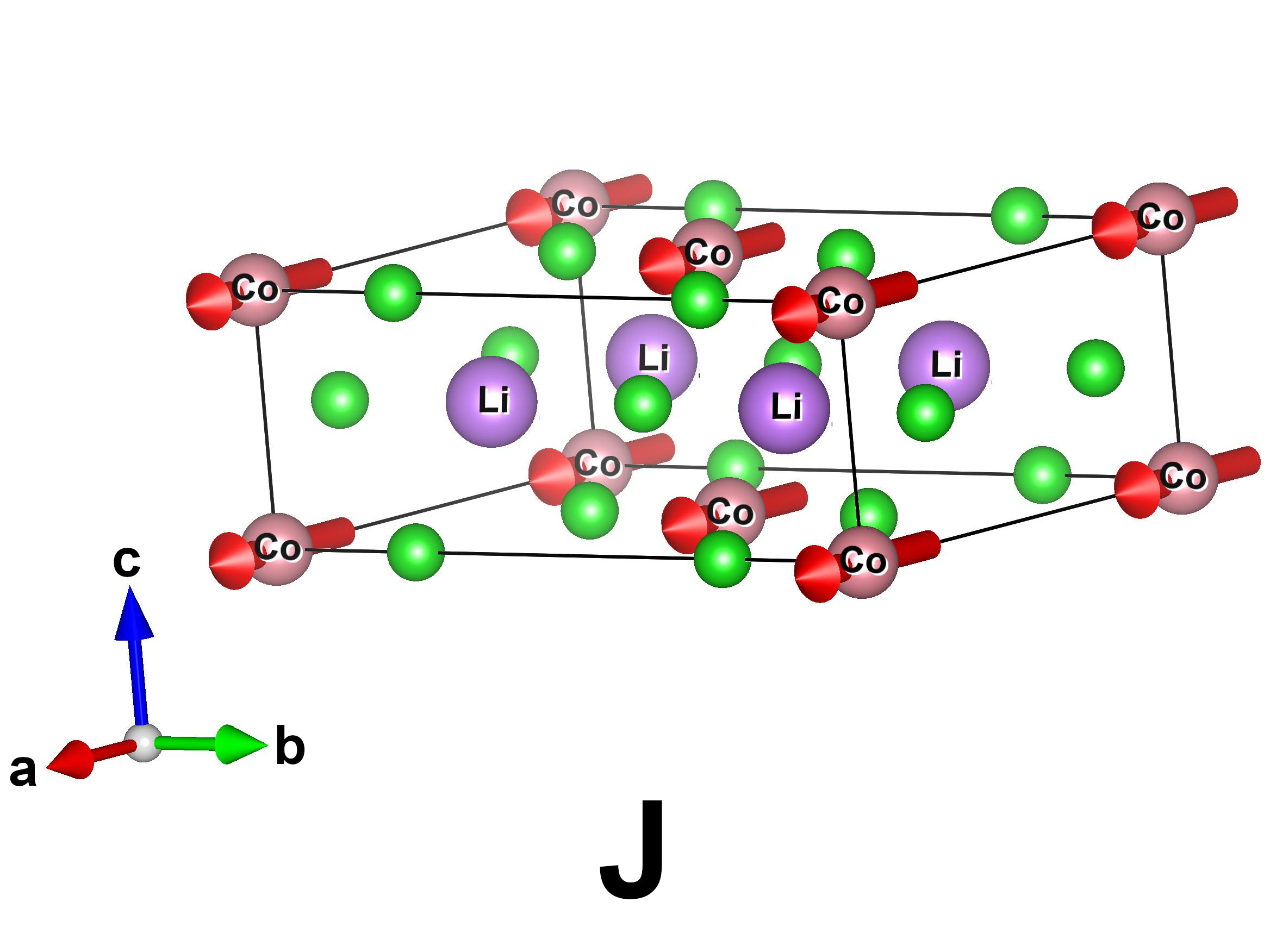}}
    \hfill
    \subfloat{\includegraphics[width=0.31\columnwidth]{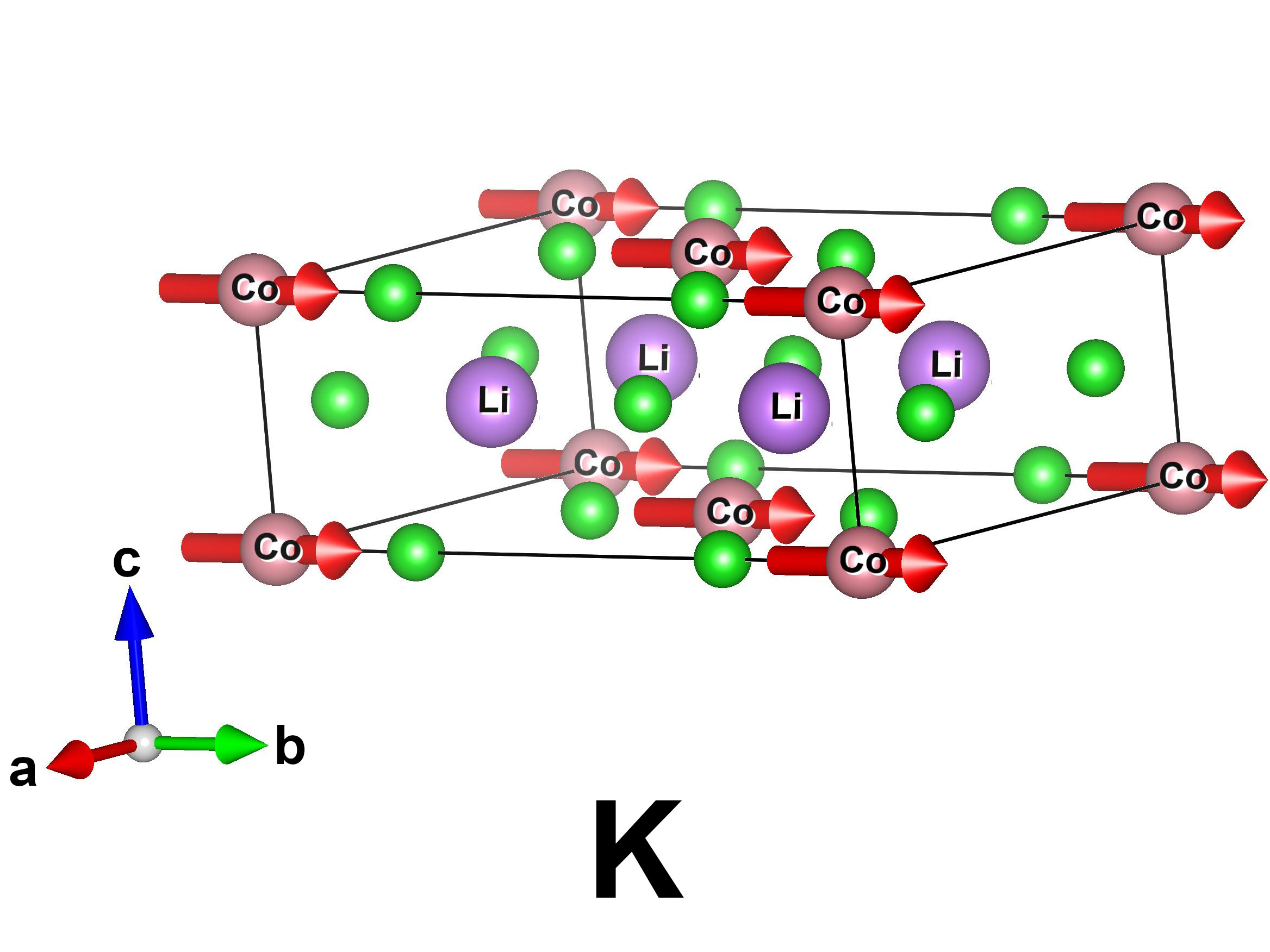}}
    \hfill
    \subfloat{\includegraphics[width=0.31\columnwidth]{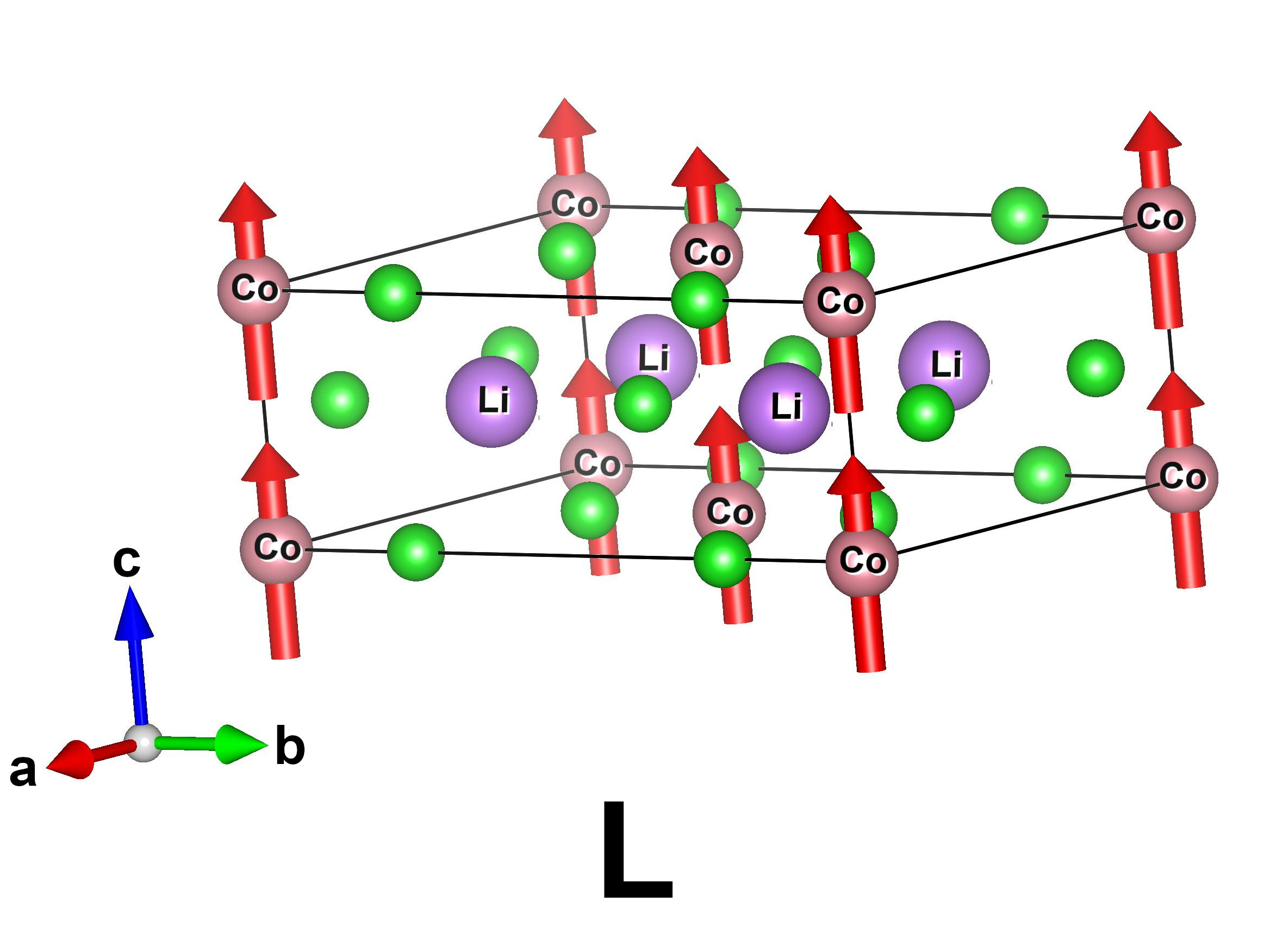}}
    \caption{The 12 proposed moment configurations used in DFT calculations are shown with the nuclear cell axes. Configurations D through I required cell doubling along the \textit{c}-axis.}
    \label{fig:DFTcells}
\end{figure}

\pagebreak
\section{Additional DFT results}
\begin{figure}[ht]
    \centering\includegraphics[width=0.5\columnwidth]{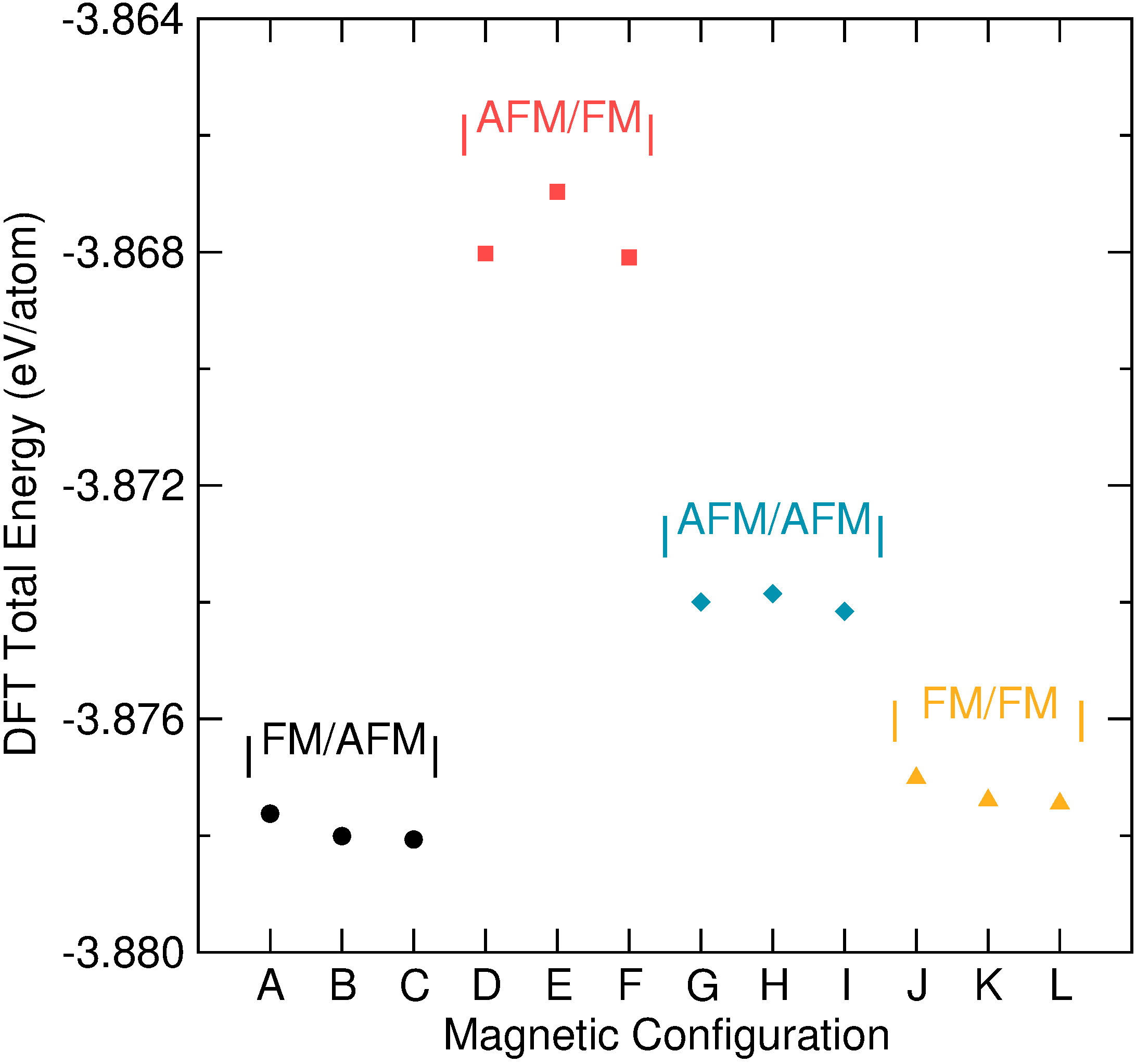}
    \caption{Unrelaxed \LCC\ cell magnetic configuration energies are shown with groups labeled by their intra-/inter-chain interactions. }
    \label{fig:dftE}
\end{figure}

DFT-calculated total energies (GGA) using an unrelaxed \LCC\ cell are presented for each magnetic configuration in Fig.~\ref{fig:dftE}. Unlike the relaxed case, the ferromagnetic set (J-L) is lower in energy than the antiferromagnetic intrachain, antiferromagnetic interchain set (G-I), and the $a$-axis is not the hard axis in the case of the D-F and G-I sets. The C configuration is now lower in energy than the B configuration by 0.062~meV/atom. The calculated cobalt moment ranges from 2.336 to 2.453~$\mu_{\mathrm{B}}$.
\clearpage
The band structure of \LCC\ was calculated with the SCAN functional for the B type configuration, which was the lowest energy configuration for the DFT-relaxed cell, following \textsc{SeeK-path}\cite{hinuma2017band} conventions. The calculation was spin-polarized and collinear without spin-orbit coupling.

\begin{figure}[ht]
    \centering\includegraphics[width=0.9\columnwidth]{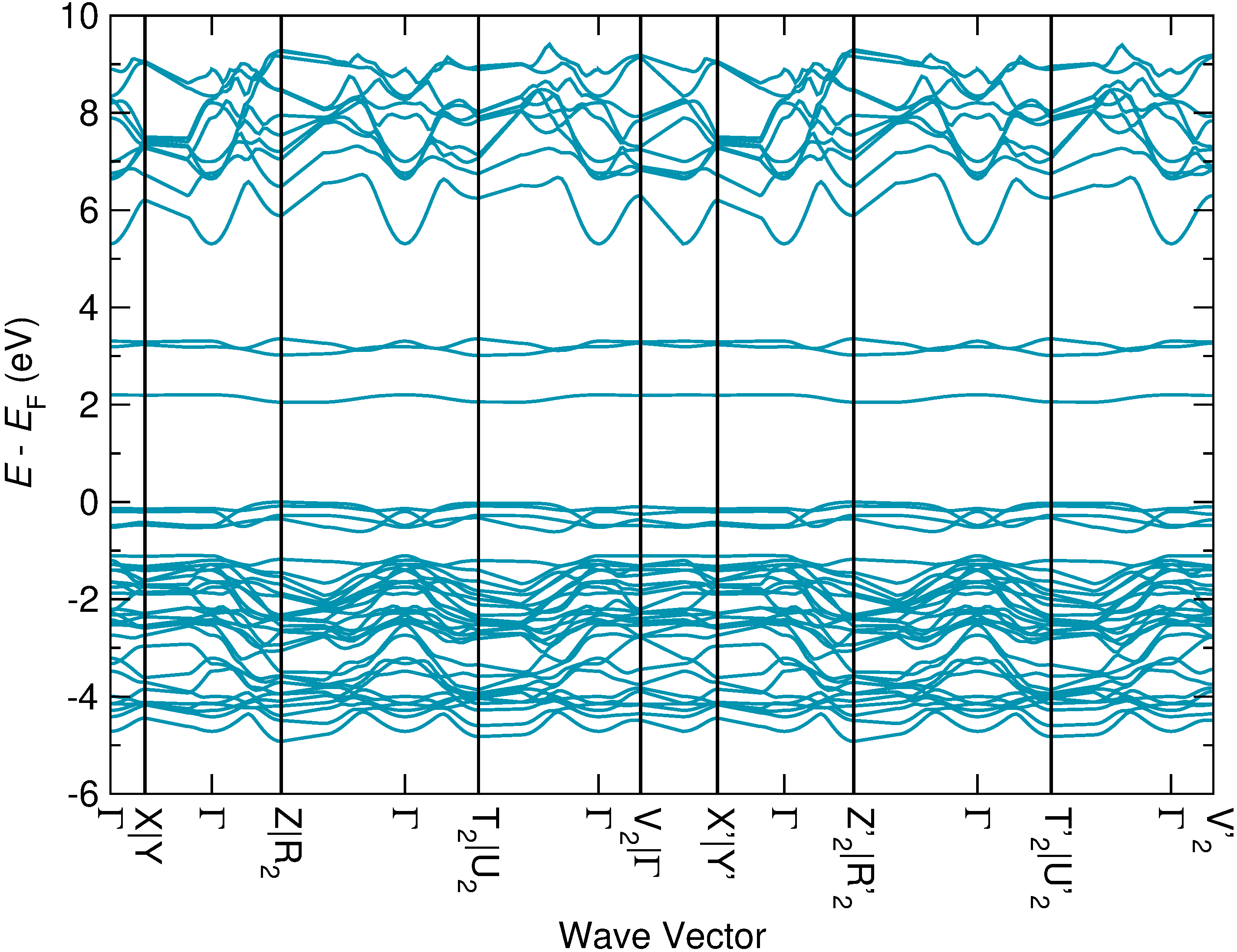}
    \caption{The \LCC\ band structure for magnetic configuration B is shown with the Fermi energy set as the energy of the highest occupied state.}
    \label{fig:dftBands}
\end{figure} 

% \FloatBarrier
\clearpage
\section{Additional zero-field powder neutron diffraction data}
\begin{figure}[h]
    \centering
 \subfloat{\includegraphics[width=0.425\columnwidth]{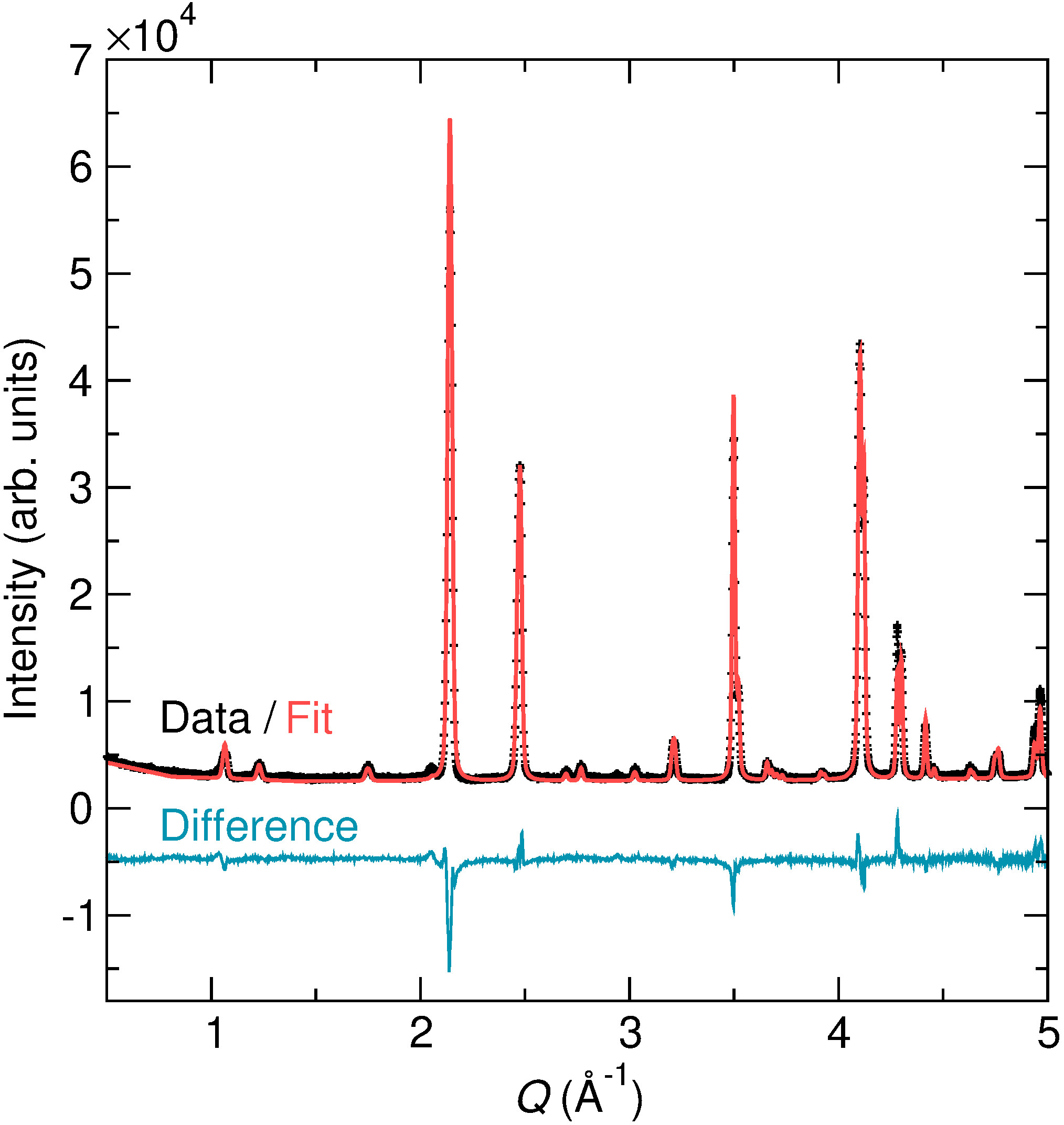}}
    \hfill
    \subfloat{\includegraphics[width=0.5\columnwidth]{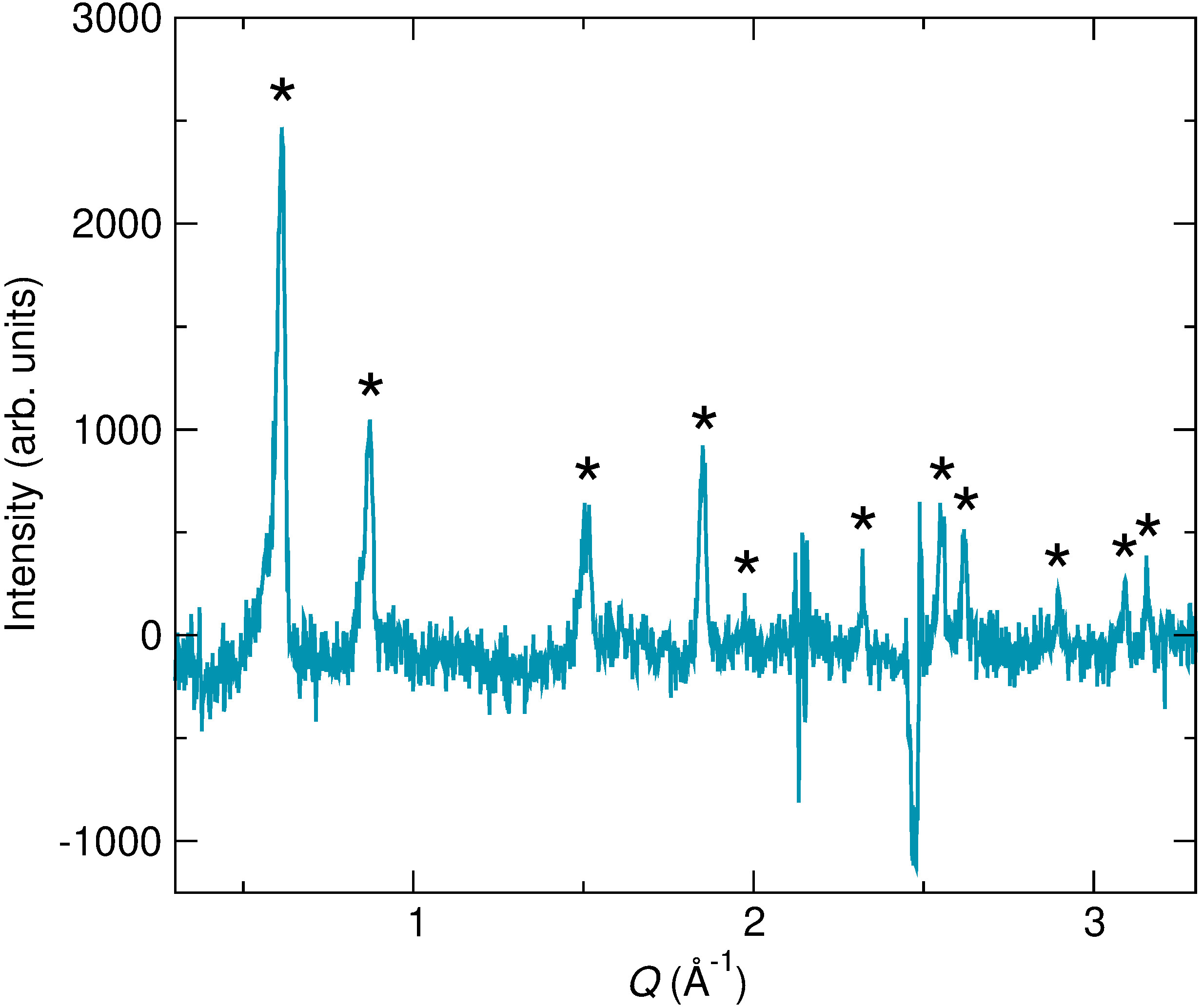}}
    \caption{(Left) We collected neutron powder diffraction data for \LCC\ at 15~K before cooling below the N\'eel temperature. The data showed no impurity peaks. (Right) The difference between the 15~K and 3.5~K neutron powder diffraction data is plotted, and the emergent magnetic peaks are marked by an asterisk.}
    \label{fig:zfnpd_xtra}
\end{figure}

\begin{table}
    \centering
    \caption{The starting lattice parameters and atomic positions for DFT relaxations came from neutron diffraction data collected at 3.5~K and zero field. The refined lattice parameters were $a$=7.1390(3), $b$=10.1899(5), $c$=3.5935(2)~\AA, and $\alpha$=$\beta$=$\gamma$=90$^{\circ}$ (space group \textit{Cmmm}). }
    \begin{tabular}{l c c c}
        \hline 
        Site & x & y & z \\
        \hline
         Li & 0.25 & 0.25 & 0.5 \\
         Co & 0 & 0 & 0 \\
         Cl1 & 0.2355(6) & 0 & 0.5 \\
         Cl2 & 0 & 0.2382(5) & 0 
    \end{tabular}
    \label{tab:my_label}
\end{table}

\pagebreak

\bibliography{Li2CoCl4.bib}